\documentclass[letterpaper, paper,10pt]{AAS}		

\usepackage{bm}
\usepackage{amsmath}
\usepackage{subfigure}
\usepackage[colorlinks=true, pdfstartview=FitV, linkcolor=black, citecolor= black, urlcolor= black]{hyperref}
\usepackage{overcite}
\usepackage{footnpag}			      	

\usepackage{algorithm}
\usepackage{algorithmic}

\PaperNumber{XX-XXX}

\begin{document}

\title{Imitation Learning for Satellite Attitude Control under Unknown Perturbations}

\author{Zhizhuo Zhang\thanks{Ph.D. Student, Mechanical and Aerospace Engineering Department, Rutgers, the State University of New Jersey, 98 Brett Rd, Piscataway, NJ 08854, United States.},
Hao Peng\thanks{Assistant Professor, Aerospace Engineering Department, Embry-Riddle Aeronautical University, 1 Aerospace Blvd, Daytona Beach, FL 32114, United States.},
Xiaoli Bai\thanks{Associate Professor, Mechanical and Aerospace Engineering Department, Rutgers, the State University of New Jersey, 98 Brett Rd, Piscataway, NJ 08854, United States.}
}

\maketitle{ }

\begin{abstract}
This paper presents a novel satellite attitude control framework that integrates Soft Actor-Critic (SAC) reinforcement learning with Generative Adversarial Imitation Learning (GAIL) to achieve robust performance under various unknown perturbations. Traditional control techniques often rely on precise system models and are sensitive to parameter uncertainties and external perturbations. To overcome these limitations, we first develop a SAC-based expert controller that demonstrates improved resilience against actuator failures, sensor noise, and attitude misalignments, outperforming our previous results in several challenging scenarios. We then use GAIL to train a learner policy that imitates the expert’s trajectories, thereby reducing training costs and improving generalization through expert demonstrations. Preliminary experiments under single and combined perturbations show that the SAC expert can rotate the antenna to a specified direction and keep the antenna orientation reliably stable in most of the listed perturbations. Additionally, the GAIL learner can imitate most of the features from the trajectories generated by the SAC expert. Comparative evaluations and ablation studies confirm the effectiveness of the SAC algorithm and reward shaping. The integration of GAIL further reduces sample complexity and demonstrates promising imitation capabilities, paving the way for more intelligent and autonomous spacecraft control systems.

\end{abstract}

\section{Introduction}


Aiming at accurately orienting and stabilizing satellites towards specific directions or targets in space, satellite attitude control is a critical aspect of spacecraft missions. Satellite attitude control methodologies primarily rely on classical control theories, including PID control\cite{bolandi2013attitude} , Quaternion Rate Feedback\cite{lim2001new} , and Model Predictive Control\cite{vedant2018dynamic} , etc. Although most of these approaches ensure a certain degree of stability and accuracy, they often struggle with handling complex dynamic constraints, and typically require accurate system models\cite{gao2020satellite} . Furthermore, traditional control algorithms are sensitive to changes in system parameters. Particularly in environments with perturbations (such as orbital perturbations, atmospheric drag, or solar radiation pressure), traditional control methods often require additional compensation strategies. Examples of such strategies include the modified framework called Compatible Performance Control (CPC)\cite{lei2023adaptive} , disturbance observers to counteract unknown external torques\cite{shi2020disturbance} , etc. These supplementary measures could inevitably increase complexity and reduce robustness. Therefore, it's necessary to design more intelligent satellite attitude control algorithms with strong adaptive ability.

Recently, reinforcement learning (RL) has garnered significant attention. RL methods enable agents to interact with their environment through trial-and-error and continuous policy optimization, achieving optimal action sequences. Consequently, RL can autonomously learn complex control strategies without relying on precise models, effectively adapting to dynamic environments and variations in system parameters. In spacecraft attitude control, the advantages of RL-based approaches when encountering external perturbations become even more pronounced. By learning optimal policies under different perturbation conditions, the controllers maintain high efficiency, stability, and robustness across dynamic disturbance environments. The existing literature on applying RL methods to attitude control problems has provided some background on this topic. Vedant et al.\cite{vedant2019reinforcement} utilized the Proximal Policy Optimization (PPO) algorithm and showed promising results in dealing with spacecraft attitude control tasks involving actuator constraints such as saturation limits and momentum exchange mechanisms, achieving improved robustness compared to traditional Quaternion Rate Feedback (QRF) controllers. Jacob et al.\cite{elkins2022bridging} benchmark two RL algorithms, PPO and Twin Delayed Deep Deterministic Policy Gradient (TD3), on a spacecraft simulation, emphasizing sample efficiency and control precision. However, neither Vedant et al. nor Jacob et al. have conducted experiments employing the SAC algorithm. Hao et al.\cite{peng2023reorient} used a Deep Deterministic Policy Gradient (DDPG)-based method to reorient satellites' antennas towards the Earth and rebuild communication in a situation where unknown attitude failures have occurred. However, the agent in this work is sensitive to the initial condition of the RL environment. Maximilian\cite{meijkamprobust} assessed and compared the implementation and performance of state-of-the-art model-free reinforcement learning algorithms in the context of post-capture attitude control for an active debris removal mission, accounting for dynamic and kinematic uncertainties. However, the trained agent performed badly in all kinds of gyroscope-related experiments listed. Another notable limitation of all conventional RL methods mentioned above is their reliance on trial-and-error interactions with the environment, which can be computationally demanding and inefficient.

To address these limitations, imitation learning (IL) methods, particularly Generative Adversarial Imitation Learning (GAIL). have recently attracted attention within many applications of robotics, including self-driving cars\cite{codevilla2018end}, surgical robots\cite{ravichandar2020recent}, and multi-agent systems\cite{le2017coordinated}. Learning from Demonstration (LfD), a specific form of IL, leverages expert demonstrations to accelerate the training of complex control policies. GAIL, an IL method inspired by generative adversarial networks (GANs), enables agents to learn policies that closely mimic expert behavior by distinguishing between agent-generated and expert-generated trajectories \cite{ho2016generative} . In satellite attitude control, the integration of GAIL offers multiple advantages. First, expert demonstrations derived from traditional controllers or pre-trained RL agents can provide valuable prior knowledge, dramatically reducing the sample complexity required for policy learning. Second, policies learned via GAIL inherently retain interpretability as the learned behaviors can be traced back to comprehensible expert trajectories. Despite these potential advantages, the application of GAIL in spacecraft attitude control remains relatively unexplored, providing fertile ground for further research. 

This paper develops a novel satellite attitude control strategy, with the combination of pre-trained RL experts and imitation learning algorithms. It adapts to various disturbances in the attitude control process and improves the robustness of agents during the task of autonomously reorienting a satellite antenna to a predefined orientation. Contributions in this paper are stated below:

1. Develop an RL agent to achieve better performance than Hao et al.'s work\cite{peng2023reorient} in more perturbation experiments by refining the reward function and using the SAC algorithm.

2. Use GAIL to improve the robustness and training efficiency of RL agents.



\section{Problem Statement}

In this section, we first elaborate on the particular attitude control problem to be handled by our method. Then, we provide our insights on the choice of the related algorithms and a brief introduction to them.

\subsection{Satellites Attitude Control with Perturbations}

We assume the satellite initially suffers from an undiagnosed failure, causing the misalignment of its antenna. The RL controller must effectively leverage onboard actuators, operating under uncertainty and without direct access to the satellite’s low-level attitude determination and control functions. The primary goal is to restore the antenna orientation accurately, thereby ensuring successful reconnection to ground-based communication signals. The performance will be assessed by measuring orientation accuracy and the duration of stable communication achieved after realignment, thereby demonstrating the capability of RL to autonomously recover spacecraft functionality under disrupted conditions. 

Particularly, in this paper, perturbations of the gyroscope, the actuator, and the attitude sensor will be involved.

\subsection{Choice of Reinforcement Learning Algorithm}

From the aspect of the application, distinguishing RL from these two aspects is helpful\cite{wiering2012reinforcement} : (1) On-policy or off-policy algorithm; (2) Online or offline training. The first distinction lies in whether previous experiences can be leveraged to update the policy and/or value networks. The PPO algorithm and DDPG algorithm are the representatives in these two directions, respectively. The second difference is more straightforward: offline training algorithms typically have access to greater storage capacity and computational resources, naturally enhancing their ability to address complex problems.

In satellite attitude control problems discussed in this paper, the state-action relationships exhibit continuous, highly nonlinear dynamics characterized by classical mechanics and Euler’s equations of rotational motion.

Although DDPG provides a deterministic policy suitable for continuous action spaces, the critic network may overestimate the Q-value, thereby misleading the actor strategy update and producing suboptimal behavior. PPO, on the other hand, addresses policy update stability through clipped surrogate objectives. However, this mechanism also limits the possibility of large-scale policy updates, which may lead to slower learning speed, especially slow convergence or stagnant performance in complex environments.

Soft Actor-Critic (SAC)\cite{haarnoja2018soft} , by contrast, incorporates entropy-based regularization into its objective function. This entropy maximization encourages a balanced exploration-exploitation trade-off through stochastic policy distributions, effectively handling the nonlinear complexity of satellite attitude dynamics. SAC's stochastic framework explicitly accounts for uncertainty and variability within the control environment, mathematically providing smoother policy updates and increased robustness against perturbations and model inaccuracies. Furthermore, SAC inherently stabilizes training by employing dual critics and a soft-update mechanism for policy networks, thus achieving mathematically stable and consistent convergence under the complex dynamical structure of satellite rotational motion.

Therefore, we have chosen the SAC algorithm as the method used in this paper. Once the online training of the SAC algorithm is finished, we will regard it as the expert and use it to produce offline trajectories in the following GAIL training.

\subsection{Choice of Imitation Learning}

In the field of imitation learning, how to effectively learn the optimal strategy from expert demonstrations has always been a core challenge. The Behavior Cloning (BC) algorithm directly imitates the actions of experts through supervised learning, which has the advantages of simple training and convenient implementation; but its main drawback is that the errors generated during the learning process will gradually accumulate, resulting in "covariate shift", making it difficult to handle states that have not been encountered during training, thereby limiting the generalization ability. In contrast, inverse reinforcement learning (IRL) indirectly learns strategies by inferring the reward function behind the expert's behavior, which can more deeply understand and generalize the expert's strategy. However, IRL usually involves complex two-stage optimization problems, has high computational costs, and is often plagued by non-unique solutions in practice.

To effectively solve the above challenges, this study chose to use the GAIL algorithm. GAIL combines the theoretical advantages of IRL and the training framework of GAN, and can directly learn a strategy distribution similar to that of an expert without explicitly modeling a reward function. This approach not only avoids the covariate drift problem in the BC algorithm but also overcomes the limitations of the high computational complexity and ambiguity of the reward function of the IRL algorithm. 

Therefore, GAIL has both training stability and computational efficiency while ensuring efficient generalization, making it the preferred algorithm for achieving the imitation learning goal in this study. We will use the GAIL algorithm to imitate the behaviors of the pre-trained expert RL agent.

\subsection{Formulation of the GAIL algorithm}

For ease of understanding, the pseudocode \ref{alg:gail} of the GAIL algorithm used in this paper is included below.

\begin{algorithm}[htb]
\caption{Generative adversarial imitation learning}
\label{alg:gail}
\begin{algorithmic}[1]
\STATE \textbf{Input:} Expert trajectories $\tau_E \sim \pi_E$, initial policy and discriminator parameters $\theta_0, w_0$.
\FOR{$i = 0, 1, 2, \dots$}
    \STATE Sample trajectories $\tau_i \sim \pi_{\theta_i}$.
    \STATE Update the discriminator parameters from $w_i$ to $w_{i+1}$ with the gradient:
    \begin{equation}
    \hat{E}_{\tau_i}\bigl[\nabla_w \log\bigl(D_w(s,a)\bigr)\bigr] 
    \;+\;
    \hat{E}_{\tau_E}\bigl[\nabla_w \log\bigl(1 - D_w(s,a)\bigr)\bigr].
    \end{equation}
    \STATE Take a policy step from $\theta_i$ to $\theta_{i+1}$ using the SAC rule with cost function 
    $log(D_{w_{i+1}}(s, a))$. Specifically, take a KL-constrained natural gradient step:
    \begin{equation}
    \hat{{E}}_{\tau_i} 
    \Bigl[
        \nabla_{\theta} \log \pi_{\theta}(a \mid s)\;Q(s,a)
    \Bigr]
    \;-\;
    \lambda \,\nabla_{\theta} H\bigl(\pi_{\theta}\bigr),
    \end{equation}
    where
    \begin{equation}
    Q(\bar{s}, \bar{a}) 
    \;=\; 
    \hat{E}_{\tau_i}
    \Bigl[
        \log\bigl(D_{w_{i+1}}(s,a)\bigr)
    \Bigr]
    \;\bigm|\; 
    s_0 = \bar{s},\; a_0 = \bar{a}.
    \end{equation}
\ENDFOR
\end{algorithmic}
\end{algorithm}


Within the GAIL architecture, the learning process involves two interacting policies: the expert's policy \(\pi_E\), which serves as a performance benchmark, and the learner's policy \(\pi_{\theta}\), tasked with imitating the expert. The learner policy aims to emulate the expert by maximizing the expected cumulative reward derived implicitly from a discriminator's evaluations:



The discriminator, parameterized by \(\omega\), is trained to differentiate state-action pairs \((s,a)\) sampled from the learner and expert policies, respectively. Specifically, the discriminator outputs probabilities that reflect its belief that a given pair originated from the learner rather than the expert. Formally, the discriminator's training objective is expressed as:
\begin{equation}
    L(\omega) = -E_{\rho_{\pi_{\theta}}}[\log D_{\omega}(s,a)] - E_{\rho_{\pi_E}}[\log(1 - D_{\omega}(s,a))],
\end{equation}
where \(\rho_{\pi_{\theta}}\) and \(\rho_{\pi_E}\) represent occupancy distributions from the learner and expert policies, respectively.

The implicit reward mechanism in GAIL is uniquely defined via the discriminator's outputs, eliminating explicit reward engineering. This reward is mathematically represented as:
\begin{equation}
    r(s,a) = -\log(D_\omega(s,a)).
\end{equation}
This reward formulation encourages the learner to generate state-action pairs indistinguishable from expert demonstrations. If the discriminator perceives a learner-generated pair as expert-like, \(D_\omega(s,a)\) approaches unity, resulting in a less negative reward. Conversely, significant deviation from expert behaviors drives \(D_\omega(s,a)\) closer to zero, resulting in harsher penalties, thus continually steering the learner towards more accurate imitation.

Consequently, the GAIL algorithm iteratively performs two optimization steps:
\begin{enumerate}
    \item \textbf{Discriminator Update:} Optimize \(D_\omega\) to enhance its classification accuracy between learner and expert-generated trajectories.
    \item \textbf{Policy Update:} Refine the policy \(\pi_{\theta}\) to maximize the discriminator-based implicit reward, progressively aligning its behavior with the expert's demonstrations.
\end{enumerate}

Through iterative adversarial updates, GAIL effectively reduces the performance gap between the learner and the expert, delivering robust imitation capabilities without explicit reward definitions.

\section{Method}

In our paper, we first conduct the training of the SAC algorithm to let the agent learn the control strategy under perturbations. Then, we regard the pre-trained SAC algorithm as the expert and let it generate trajectories $\tau_E$. Another SAC agent, which we regard as the learner, will learn from these trajectories and produce similar trajectories $\tau_l$. Apart from that, the discriminator with an MLP-based neural network will also be trained to distinguish the difference between trajectories generated by the expert and the learner.

\subsection{Kinematics and Dynamics Simulation of the Capstone Satellite}

The quaternion $\mathbf{q} = [q_1, q_2, q_3, q_4]^\top$ describes the satellite attitude. The vector $\boldsymbol{\omega}$ expresses the angular velocity in the principal body frame. The continuous kinematic equation\cite{crassidis2004optimal} is given by
\begin{equation}
    \label{eq:q_dot}
    \dot{\mathbf{q}} = \frac{1}{2}
    \begin{bmatrix}
        -[\boldsymbol{\omega} \times] & \boldsymbol{\omega} \\
        -\boldsymbol{\omega}^\top & 0
    \end{bmatrix} \mathbf{q} 
\end{equation}
where $[\boldsymbol{\omega} \times]$ is the skew-symmetric matrix representing the cross product of $\boldsymbol{\omega}$.

The attitude dynamic equation \cite{crassidis2004optimal} is expressed in the body principal frame as
\begin{equation}
    \label{eq:w_dot}
    \dot{\boldsymbol{\omega}} = J^{-1} \left( -[\boldsymbol{\omega} \times] J \boldsymbol{\omega} + \mathbf{M} \right)
\end{equation}
in which $\boldsymbol{\omega} = [\omega_x, \omega_y, \omega_z]^\top$ is the angular velocity, and $\mathbf{M}$ represents the total external torque acting on the satellite. For this study, we assume that $\mathbf{M} = [M_x, M_y, M_z]$ contains the torques applied along the three body-fixed axes. While real-world satellites may utilize more advanced actuation mechanisms such as thrusters, reaction wheels, or control moment gyroscopes (CMGs), we focus only on torque inputs for simplicity and clarity.

To simulate the system dynamics, we numerically solve Eqs. \ref{eq:q_dot} and \ref{eq:w_dot} using the MuJoCo Physical Engine \footnote{\url{https://mujoco.readthedocs.io/en/stable/overview.html}}. Quaternion representation, as shown in Eqs. \ref{eq:q_dot}, is sensitive to numerical drift. MuJoCo's stable quaternion normalization ensures accurate and drift-free attitude representation throughout extended simulations. Additionally, with its comprehensive API, MuJoCo can use imported MJCF mesh to implement detailed customization of the inertia tensor (\(\mathbf{J}\)) and external forces (\(\mathbf{M}\)), closely matching real-world satellite specifications. The time step is set to 0.01 seconds for integration. Torques are updated every 10 integration steps to strike a balance between computational efficiency and simulation accuracy.

\begin{figure}[htb]
    \centering
    \includegraphics[width=0.5\linewidth]{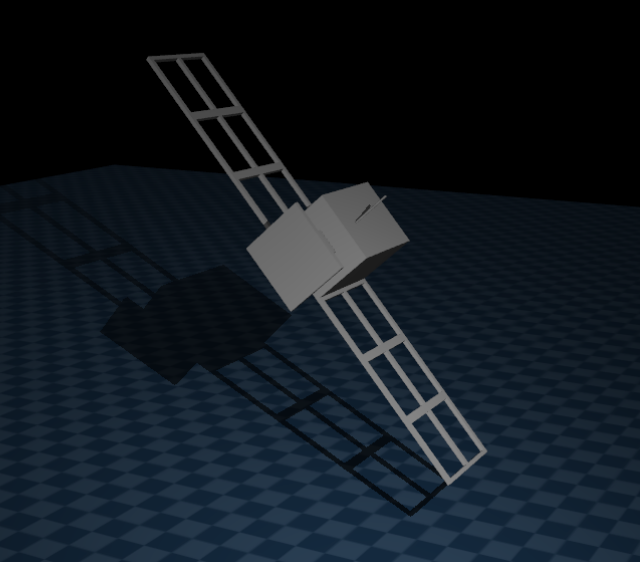}
    \caption{Visualization for the CAPSTONE Satellite in MuJoCo}
    \label{fig:CapstoneMuJoCo}
\end{figure}

Our satellite model is built based on the CubeSat mission CAPSTONE \cite{cheetham2021cislunar}, as shown in Figure \ref{fig:CapstoneMuJoCo}. Based on the fact that the weight of CAPSTONE is around 25~kg and the dimension is around $34 \times 34 \times 64$~cm (unfolded), the rotational inertia $J$ is approximated as
\begin{equation}
    J = \mathrm{diag}(J_x, J_y, J_z) = \mathrm{diag}(0.482, 1.094, 1.100)\ \mathrm{kg} \cdot \mathrm{m}^2.
\end{equation}

Where the rotational inertia is identical to that used in Hao et al.'s work\cite{peng2023reorient} .

We note that $J_z > J_y$ is assumed to be slightly larger to take the expanded solar panels into consideration. The control torque $\mathbf{M}$ is restricted to $[-0.1, 0.1]$~Nm for each principal axis. During the simulation, $\mathbf{M}$ is commanded by the learned RL policy.

\subsection{Perturbations Parameters}

Based on Hao et al.'s work\cite{peng2023reorient} and Maximilian's work\cite{meijkamprobust} , we design several kinds of perturbations in this paper. Due to computational limitations, two kinds of perturbations are applied at the same time.

\subsubsection{Gyroscope Noise} 

To account for sensor imperfections, we introduce a noise component to the angular velocity measurements, simulating gyroscope noise. This reflects the practical scenario where attitude rates are inferred from noisy sensor outputs. The model assumes that the measurement is corrupted by Gaussian white noise, superimposed on the true angular velocity signal:
\begin{equation}
    \tilde{\boldsymbol{\omega}} = \boldsymbol{\omega} + \boldsymbol{\varepsilon}
\end{equation}
where $\boldsymbol{\varepsilon}_i \sim \mathcal{N}(0, \sigma_\omega^2),\ i = 1,2,3$ is a white noise term sampled independently at each time step.

In the ideal scenario (unperturbed), the noise variance is set to zero, i.e., $\sigma_\omega = 0^\circ/s$. In contrast, for simulating realistic sensor behavior, we consider a noise standard deviation of $\sigma_\eta = 0.1^\circ/s$r, as reported in experimental studies of gyroscopic noise \cite{thienel2003coupled} . This value provides a credible approximation for on-board rate sensor uncertainties and enhances the fidelity of the simulation.

\subsubsection{Constant Gyroscope Bias} 

In addition to white noise, gyroscope measurements may exhibit a persistent bias arising from hardware imperfections or calibration inaccuracies. This bias is modeled as an additive constant vector in the angular velocity readings:
\begin{equation}
    \label{eq:gyro_bias}
    \tilde{\boldsymbol{\omega}} = \boldsymbol{\omega} + \mathbf{b}
\end{equation}
where $\mathbf{b}$ denotes a time-invariant bias vector. The magnitude of this bias is selected based on values commonly reported for real-world gyroscopes. In this work, we model the bias vector as $\mathbf{b} = [b_1,\ b_2,\ b_3]^\top$, with each component drawn from a Gaussian distribution: $b_i \sim \mathcal{N}(0, \sigma_b^2)$, for $i = 1, 2, 3$.

In the nominal (bias-free) setting, we use $\sigma_b = 0^\circ/s$, resulting in $\mathbf{b} = \mathbf{0}$. For the perturbed case, the standard deviation is set to $\sigma_b = 0.1^\circ/s$. Once initialized, this bias remains constant throughout the entire simulation episode.

\subsubsection{Gyroscope Drift} 

In this formulation, we extend the constant bias model (Eq. (\ref{eq:gyro_bias})) to account for time-varying drift using a stochastic process. Specifically, the gyroscope bias evolves according to a random walk, described by the stochastic differential equation:
\begin{equation}
    db_t = \sigma_b\, dW_t
\end{equation}
where $W_t$ denotes a Wiener process (Brownian motion) with increments defined by $W_t - W_{t-1} \sim \mathcal{N}(0, \Delta t)$. 

Because the bias originates from sensor drift, it is assumed to update at the same temporal resolution as the agent’s sampling frequency. In our case, this corresponds to a time step of $\Delta t = 0.1$ seconds. The parameter $\sigma_b$ determines the intensity (variance rate) of the random walk.

We initialize the bias with $b(0) = 0$. In a noise-free scenario, $\sigma_b = 0$, which implies $b(t) = 0$ for all $t$. To simulate realistic sensor behavior, we adopt a perturbation level of $\sigma_b = 0.01, 0.001, 0.0001$.

\subsubsection{Torque Failure}

Specifically, we examine scenarios where the control torques applied along the principal body axes are partially or completely corrupted.

Although the RL policy continues to produce torque commands $\mathbf{M}$ under the assumption of perfect actuation, the real-world actuators may deliver altered versions of these torques due to hardware faults or unpredictable disturbances. For simulation purposes, we define a constant deterministic scaling factor $\gamma \in (0, 1)$ that determines the reliable portion of each commanded torque. The remaining $(1 - \gamma)$ is modeled as a random component.

As an example, consider a case where the torque along the $x$-axis is degraded. The applied torque $M'_x$ is then defined as 
\begin{equation}
    M'_x = \xi_x M_x
\end{equation}
where $\xi_x$ is sampled from a uniform distribution over $[\gamma, 1]$. 

In more severe failure cases, the torque along an entire axis might be nullified. For instance, if $M_x$ completely fails, the torque vector becomes:
\begin{equation}
    \mathbf{M}' = [0,\ M_y,\ M_z]^\top.
\end{equation}
It is important to note that while randomly scaled torques still offer some influence over the system dynamics, complete failure of an axis reduces controllability and can significantly impact the satellite's ability to maintain or recover a desired attitude.

\subsubsection{Attitude Misalignment}

Another potential failure mode concerns the satellite's attitude determination subsystem. Such failure can arise from sudden mechanical disruptions—such as collisions or explosions—that misalign attitude sensors or shift the antenna structure, ultimately causing incorrect directional readings.

In such scenarios, the control system may operate based on faulty attitude data, leading to poor or even failed stabilization. While advanced filtering techniques can be developed to estimate and correct such misalignments, they may not always work reliably in unpredictable conditions. 

To simplify the analysis, this study focuses on a case with a fixed misalignment in the reported attitude. Specifically, we model this misalignment using a set of constant Euler angle offsets following the 3-2-1 sequence: $(\phi, \theta, \psi)$. These angles are converted to a quaternion representation denoted as $\delta \mathbf{q}$.

In the simulation, the agent does not receive the true attitude $\mathbf{q}$ but instead observes a perturbed version $\mathbf{q}'$, computed as:
\begin{equation}
    \mathbf{q}' = \delta \mathbf{q} \otimes \mathbf{q}
\end{equation}
where $\otimes$ denotes quaternion multiplication. This means the agent is always acting under the influence of a consistent bias in attitude estimation.

\subsection{SAC Algorithm Configuration}

The reward function is the most important part of the RL agent. The design of the reward function directly depends on the goal of the task and has a great impact on the final test performance. In this paper, the reward function designed for the training of the SAC expert is stated in Eq. (\ref{eq:reward})


\begin{equation}
\label{eq:reward}
\renewcommand{\arraystretch}{1.8} 
\left\{
\begin{array}{ll}
\text{attitude reward} = \exp\left(\frac{-\phi_{t}}{0.14 \cdot 2\pi} \right) \cdot s_1 \\ 
\text{control reward} = -\frac{\left\| M_t \right\| }{\left\| M_{\max} \right\|} \cdot s_2 \\
\text{worse penalty} = -1 \cdot s_3 & \text{if } \phi_t > \phi_{t-1} \\
\text{stay bonus} = 1 \cdot s_4 & \text{if } \phi_t \leq 5^\circ \\
\text{out-of-bound penalty} = -1 \cdot s_5 & \text{if terminated } (\omega > 10^\circ/\text{s}) \\
\end{array}
\right.
\end{equation}

where, $s_1, s_2, s_3, s_4, s_5$ means scaler for each item separately. Based on Maximilian's work\cite{meijkamprobust} and Jacob's work\cite{elkins2022bridging}, we set the default value of them as $s_1 = 1, s_2 = 0.5, s_3 = 1.0, s_4 = 9, s_5 = 500$. Specifically, in Jacob's work, $s_5$ is 25, we adjust it in our implementation because this reward term is discrete and the proportion of it in the total reward of one episode will significantly affect the performance. Considering the total simulation time per episode in our implementation is different from that in Jacob's work, this adjustment is reasonable. Other hyperparameters are listed in Appendix A.

\subsubsection{Attitude Reward}:
\[
\exp\left(-\frac{\phi_t}{0.14 \cdot 2\pi}\right) \cdot s_1
\]
This term incentivizes the agent to minimize the attitude deviation \( \phi_t \). The exponential form ensures a rapid decay of reward with increasing deviation, promoting precise alignment with the target orientation.

\subsubsection{Control Reward}:
\[
-\left\|\frac{M_t}{M_{\text{max}}}\right\| \cdot s_2
\]
This term penalizes large control torques \( M_t \), normalized by the maximum allowable torque \( M_{\text{max}} \), thereby encouraging energy-efficient control actions.

\subsubsection{Worse Penalty}:
\[
-1 \cdot s_3 \quad \text{if } \phi_t > \phi_{t-1}
\]
This penalty is applied when the attitude deviation worsens compared to the previous timestep, discouraging regressions in control accuracy.

\subsubsection{Stay Bonus}:
\[
1 \cdot s_4 \quad \text{if } \phi_t \leq 5^\circ
\]
A bonus reward is awarded when the deviation remains within an acceptable bound of \( 5^\circ \), promoting long-term stability in the controlled attitude.

\subsubsection{Out-of-Bound Penalty}:
\[
-1 \cdot s_5 \quad \text{if terminated} \left( \omega > 10^\circ/\text{s} \right)
\]
A severe penalty is applied if the angular velocity \( \omega \) exceeds the safety threshold of \( 10^\circ/\text{s} \), leading to episode termination. This deters risky or unstable maneuvers.

\subsection{GAIL Algorithm Configuration}


We regard the pre-trained SAC algorithm as the expert and let it generate trajectories $\tau_E$. Another SAC agent, which we regard as the learner, will learn from these trajectories and produce similar trajectories $\tau_l$. Apart from that, the discriminator with an MLP-based neural network will also be trained to distinguish the difference between trajectories generated by the expert and the learner. For each perturbation case, a dedicated learner is trained to imitate the behavior of the specific expert.

\subsection{Quantitative Evaluation Metrics}

Hao et al. \cite{peng2023reorient} only use the $10^\circ$ threshold to evaluate whether the antenna of the satellite has been reoriented towards the desired direction by the trained RL model. This metric is only suitable for rough result evaluation. Therefore, we introduce several quantitative metrics to evaluate the advantages and disadvantages of different methods more accurately.

\begin{enumerate}
  \item Duty Cycle
  \[
    P_{\mathrm{in}}
      = \frac{1}{T} 
        \int_{0}^{T} 
          \mathbf{1}\bigl[\theta(t)\le \theta_{\mathrm{thr}}\bigr]
        \,\mathrm{d}t,
  \]

  \item Root mean square (RMS) Error
  \[
    \theta_{\mathrm{RMS}}
      = \sqrt{\frac{1}{T}\int_{0}^{T}\theta^2(t)\,\mathrm{d}t}\,.
  \]

  \item Maximum Error and Minimum Error
  \[
    \theta_{\max}
      = \max_{t\in[0,T]}\,\theta(t)\, \quad
    \theta_{\min}
      = \min_{t\in[0,T]}\,\theta(t)\,
  \]
  
\end{enumerate}

where $\theta(t)$ is the angular difference during the evaluation, $\theta_{\mathrm{thr}}=10^\circ$, and $\mathbf{1}[\cdot]$ is the indicator function. $T$ means the maximum time step during the evaluation. It could be $2500s$ or $5000s$, depending on different experiments. The higher duty cycle value means more time the antenna of the satellite is within the $10^\circ$ threshold. The lower RMS error value means less fluctuation of the angular difference curve. The lower maximum error and minimum error value mean lower outliers and higher theoretical optimal performance of the trained RL agents.

In the following evaluations, the value of these metrics is the average value obtained from all six initial states.

\section{Experiments and Results}

\subsection{SAC Expert Algorithm}

\begin{table}[htb]
    \fontsize{10}{10}\selectfont
    \caption{List of Perturbation Experiments}
    \label{tab:PerExperiment}
        \centering 
    \begin{tabular}{c | r} 
        \hline 
        Index & Description \\
        \hline 
        1 & no\_control\_error + no\_misalignment + no\_gyro\_error (baseline)\\
        2 & x\_fail + no\_misalignment + no\_gyro\_error\\
        3 & y\_fail + no\_misalignment + no\_gyro\_error\\
        4 & z\_fail + no\_misalignment + no\_gyro\_error\\
        5 & x\_fail + misalignment + no\_gyro\_error\\
        6 & y\_fail + misalignment + no\_gyro\_error\\
        7 & z\_fail + misalignment + no\_gyro\_error\\
        8 & xyz\_noise + misalignment + no\_gyro\_error\\
        9 & no\_control\_error + no\_misalignment + gyro\_constant\\
        10 & no\_control\_error + no\_misalignment + gyro\_noise\\
        11 & no\_control\_error + no\_misalignment + gyro\_drift\\
        12 & xy\_fail + no\_misalignment + no\_gyro\_error\\
        13 & yz\_fail + no\_misalignment + no\_gyro\_error\\
        14 & xz\_fail + no\_misalignment + no\_gyro\_error\\
        \hline
    \end{tabular}
\end{table}

In this section, the list of experiments of several perturbations is reported in Table \ref{tab:PerExperiment}. For each case, a dedicated RL agent is trained, and then the testing is performed to evaluate the performance of the trained RL agent using the six randomly generated initial states. All the RL agents have used the same reward function in Eq. (\ref{eq:reward}).

\subsubsection{Performance in the Baseline Experiment without any Error}

The situation without any control error is first examined as the baseline. For this experiment, RL agents have run 200,0000 iterations to get the final trained models.

\begin{figure}[htb]
    \centering
    \includegraphics[width=0.7\linewidth]{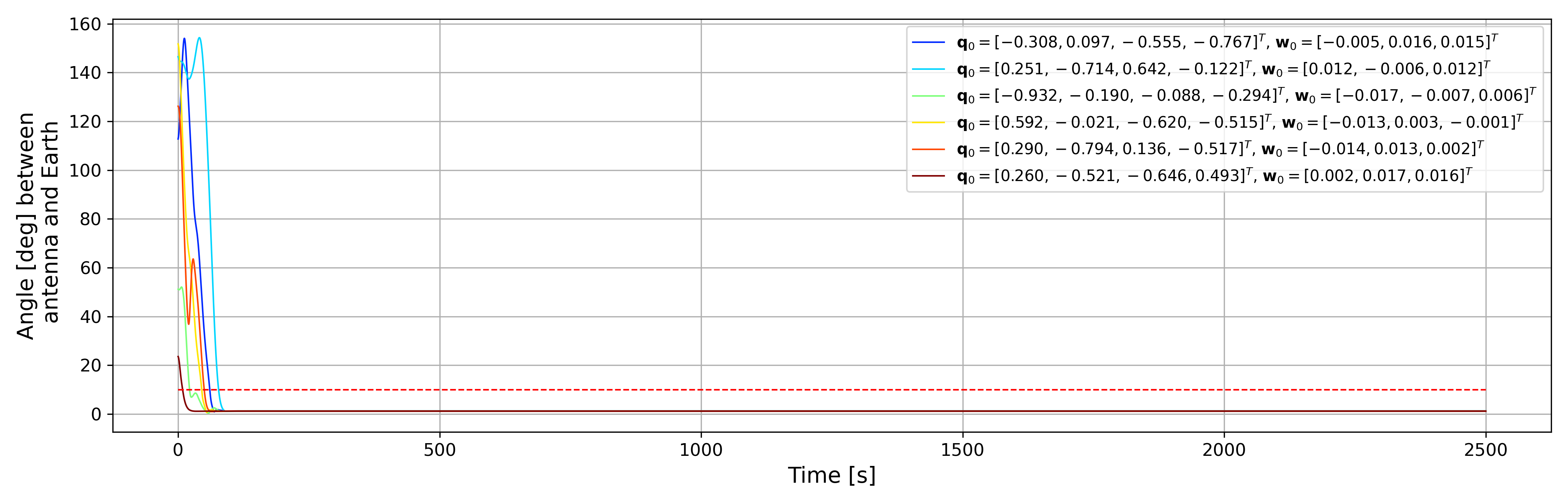}
    \caption{The evaluation of the angle difference $\alpha$ in various initial conditions (Experiment: Baseline)}
    \label{fig:baseline}
\end{figure}

Figure \ref{fig:baseline} shows the angle difference between the antenna and the target direction. The red horizontal line in the figure represents $\alpha_{max} = 10^\circ$, i.e., the half-beamwidth angle of the antenna. When $\alpha$ is within this threshold, we consider the satellite has successfully reached the target direction. As shown in Figure \ref{fig:baseline}, the learned RL policy has clearly stabilized quickly within the threshold.

\subsubsection{Performance in Experiments with Single Torque Failure}

In this section, we demonstrate the performance of the RL agent under single torque failures, which are unknown to the RL agent. For all the cases in this subsection, RL agents have run 400,0000 iterations to get the final trained models.

\begin{figure}[htb]
    \centering
    \subfigure[$M_x$ Failure Experiment \label{fig:singlefailure-a}]{
        \includegraphics[width=0.7\textwidth]{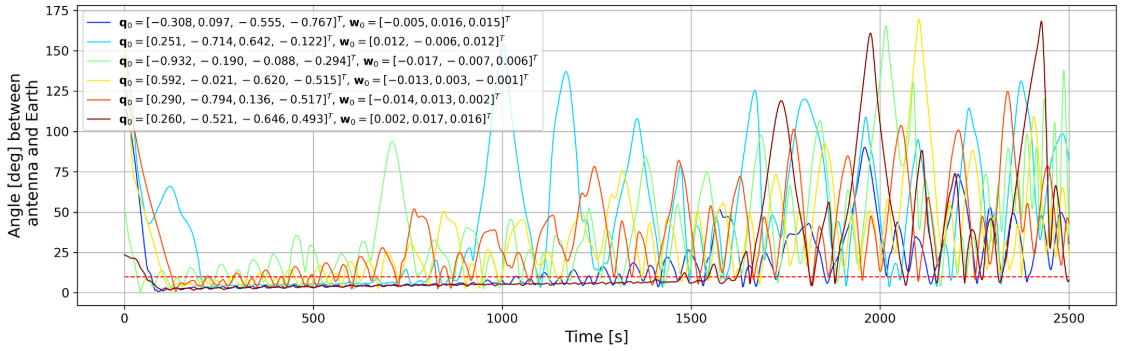}
    }
    \\ 
    \subfigure[$M_y$ Failure Experiment \label{fig:singlefailure-b}]{
        \includegraphics[width=0.7\textwidth]{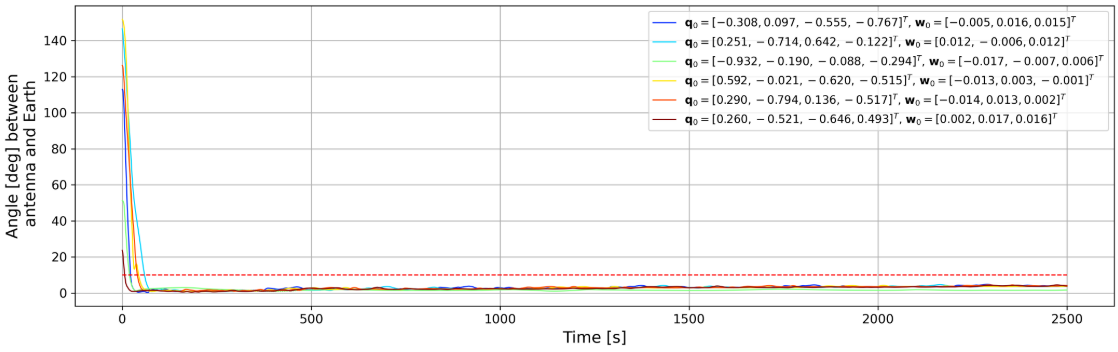}
    }
    \\ 
    \subfigure[$M_z$ Failure Experiment \label{fig:singlefailure-c}]{
        \includegraphics[width=0.7\textwidth]{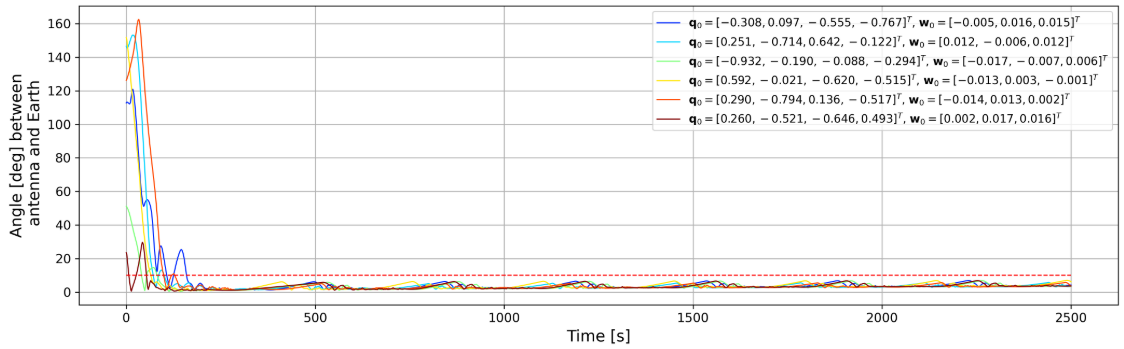}
    }
    \caption{The evaluation of the angle difference $\alpha$ in various initial conditions (Experiment: Single Torque Failure)}
    \label{fig:singlefailure}
    
\end{figure}


In Figure \ref{fig:singlefailure}, it can be observed that the performance of the $M_x$ Failure Experiment is worse than the other two experiments. Considering the rotational inertia $J$ of the satellite is different with respect to each axis, this phenomenon is expected.

There is no oscillation happening during our $M_y$ and $M_z$ Failure Experiments (Figure \ref{fig:singlefailure-b} and Figure \ref{fig:singlefailure-c}). However, we can see the results of the same experiment in Hao et al.'s work\cite{peng2023reorient} , which are shown in Figure \ref{fig:singlefailure(hao)} in Appendix B. In these previous results, Figure \ref{fig:singlefailure(hao)-b} and Figure \ref{fig:singlefailure(hao)-c} show that the RL agent failed to control the angle difference within the threshold in some cases, leading to some unexpected oscillations. Additionally, even though our trained RL agent cannot control the angle error within the threshold for all initial states in $M_x$ Failure Experiment, the angles between the antenna and the target direction are kept stable in the first half of the process for most of the initial states, while in Hao's result shown in Figure \ref{fig:singlefailure(hao)-a}, the agent failed in the first half.

Therefore, we can conclude that the RL agent we trained has indeed learned some knowledge about the current attitude failure status and achieved better performance than Hao et al.'s work\cite{peng2023reorient}.

\subsubsection{Performance in Experiments with Torque Noises and Attitude Misalignment}

In this section, we demonstrate the performance of the RL agent under three-axis torque noises and constant attitude misalignment, which are unknown to the RL agent. The particular Euler angles are arbitrarily chosen as $[15^\circ, 18^\circ, 21^\circ]$, which is a big misalignment to examine the performance of the trained RL policy. For this experiment, RL agents have run 100,0000 iterations to get the final trained models.

\begin{figure}[htb]
    \centering
    \includegraphics[width=0.7\linewidth]{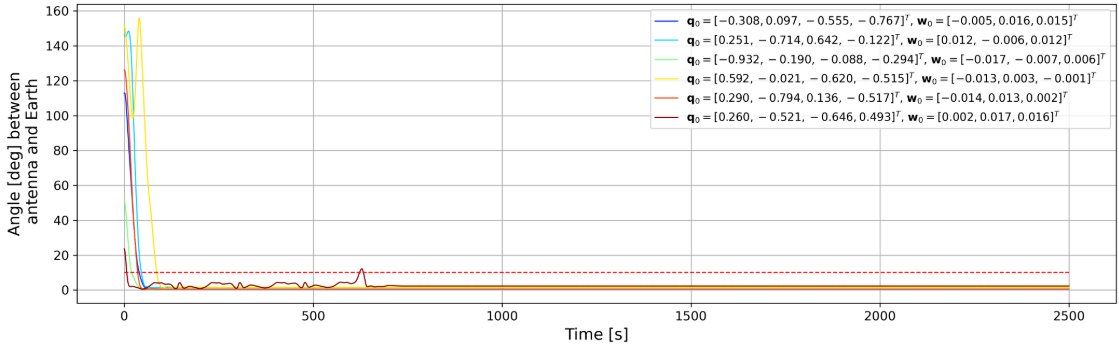}
    \caption{The evaluation of the angle difference $\alpha$ in various initial conditions (Experiment: Torque Noises and Attitude Misalignment)}
    \label{fig:torquenoise+misalignment}
\end{figure}

In Figure \ref{fig:torquenoise+misalignment}, we find that the trained RL agent can make the satellite reach the target direction regardless of the misalignment. More specifically, the RL agent learns about how to guide a state using an action with noise and figure out the misalignment during the task. Similarly, Hao et al. \cite{peng2023reorient} have also obtained a similar result in Appendix B.

\subsubsection{Performance in Experiments with Single Torque Failure and Attitude Misalignment}

In this section, we make the torque error experiment more difficult, which means we demonstrate the performance of the RL agent under single torque failures and constant attitude misalignment, which are unknown to the RL agent. The setting of misalignment is the same as that in the last section. For all the cases in this subsection, RL agents have run 400,0000 iterations to get the final trained models.

\begin{figure}[htb]
    \centering
    \subfigure[$M_x$ Failure and Attitude Misalignment Experiment \label{fig:singlefailure+misalignment-a}]{
        \includegraphics[width=0.7\textwidth]{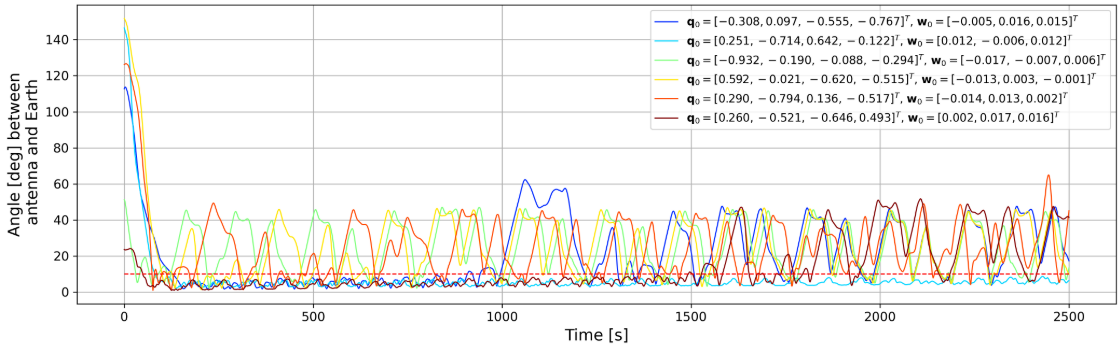}
    }
    \\ 
    \subfigure[$M_y$ Failure and Attitude Misalignment Experiment \label{fig:singlefailure+misalignment-b}]{
        \includegraphics[width=0.7\textwidth]{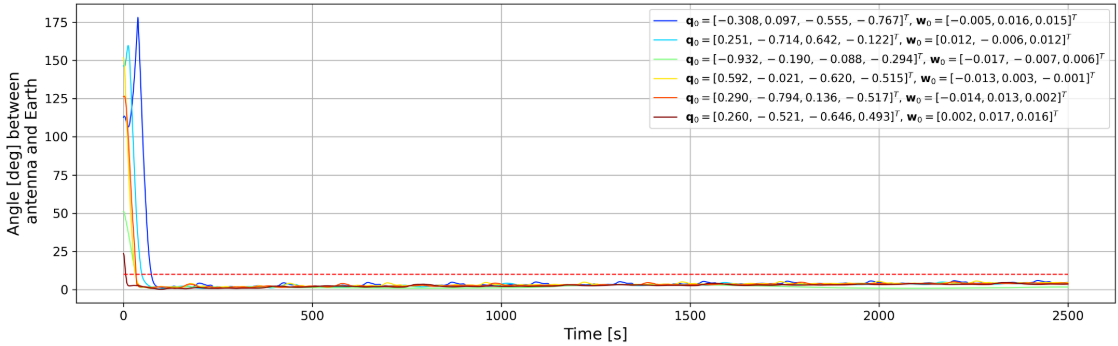}
    }
    \\ 
    \subfigure[$M_z$ Failure and Attitude Misalignment Experiment \label{fig:singlefailure+misalignment-c}]{
        \includegraphics[width=0.7\textwidth]{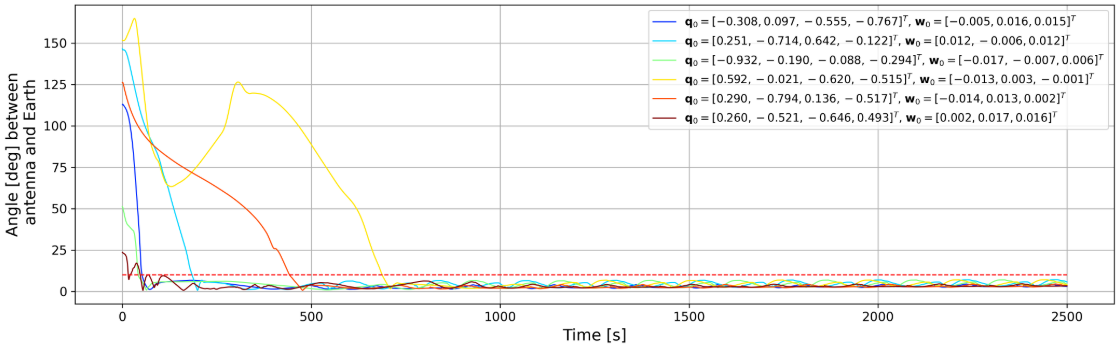}
    }
    \caption{The evaluation of the angle difference $\alpha$ in various initial conditions (Experiment: Single Torque Failure and Constant Attitude Misalignment)}
    \label{fig:singlefailure+misalignment}
\end{figure}

In Figure \ref{fig:singlefailure+misalignment}, we find that the trained RL agent can handle most cases. Compared with the results shown in Figure \ref{fig:singlefailure}, it is obvious that the RL agent can overcome the effect of misalignment and get similar performance to experiments without the constant misalignment errors. Even if the agent fails to stabilize the angle differences of all initial states within the threshold in Figure \ref{fig:singlefailure+misalignment-a}, the angle differences show that the satellites are not out of control, and they can periodically point in the target direction.

For comparison, even though Hao et al.\cite{peng2023reorient} got the similar results in Figure \ref{fig:singlefailure+misalignment(hao)-b} and Figure \ref{fig:singlefailure+misalignment(hao)-c}, the average time within the threshold in our results shown in Figure \ref{fig:singlefailure+misalignment-b} and Figure \ref{fig:singlefailure+misalignment-c} is longer than that in previous results. Additionally, the blue line and the brown line of their result in Figure\ref{fig:singlefailure+misalignment-a} show that the satellite has entered an uncontrolled rapid rotation. This phenomenon indicates that the robustness of their model needs to be improved.

\subsubsection{Performance in Experiments with Two Torque Failures}

In this section, we apply torque failures along two axes to the satellite. For all the cases in this subsection, RL agents have run 400,0000 iterations to get the final trained models.

\begin{figure}[htb]
    \centering
    \subfigure[$M_x$ and $M_y$ Failure Experiment \label{fig:twofailure-a}]{
        \includegraphics[width=0.7\textwidth]{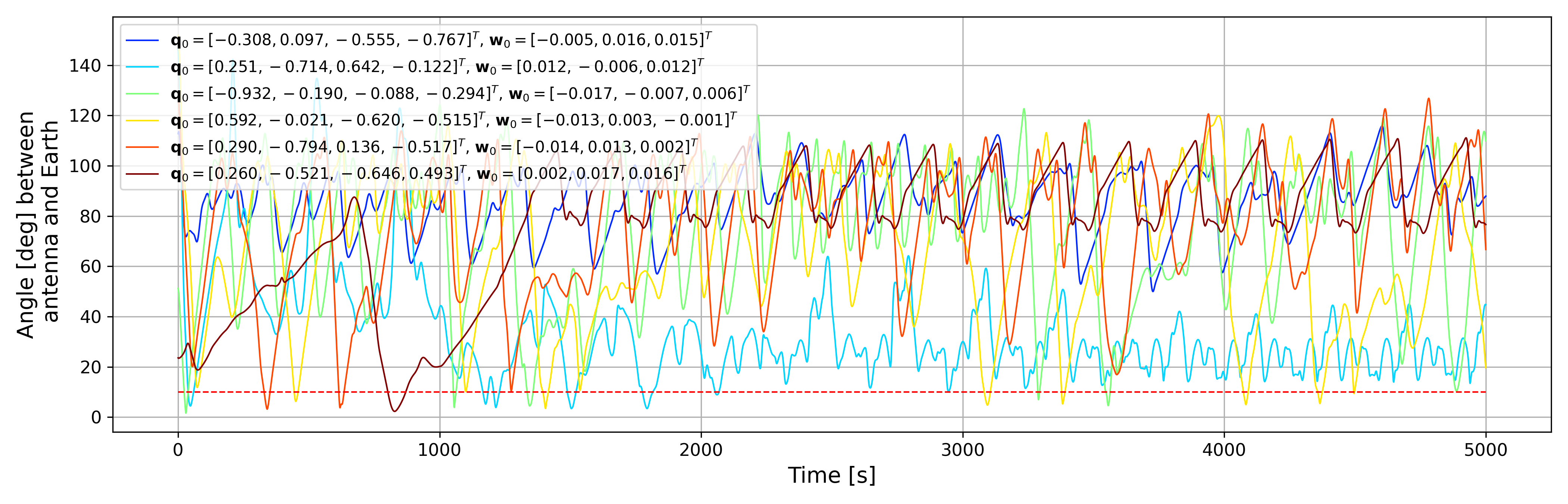}
    }
    \\ 
    \subfigure[$M_y$ and $M_z$ Failure Experiment \label{fig:twofailure-b}]{
        \includegraphics[width=0.7\textwidth]{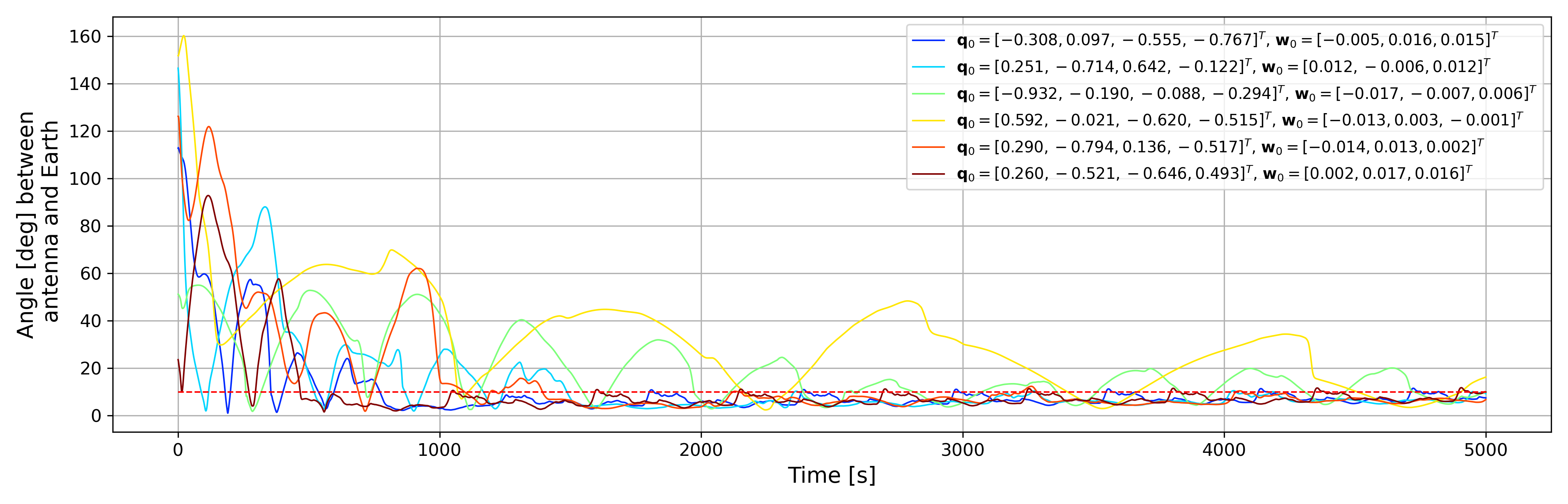}
    }
    \\ 
    \subfigure[$M_x$ and $M_z$ Failure Experiment \label{fig:twofailure-c}]{
        \includegraphics[width=0.7\textwidth]{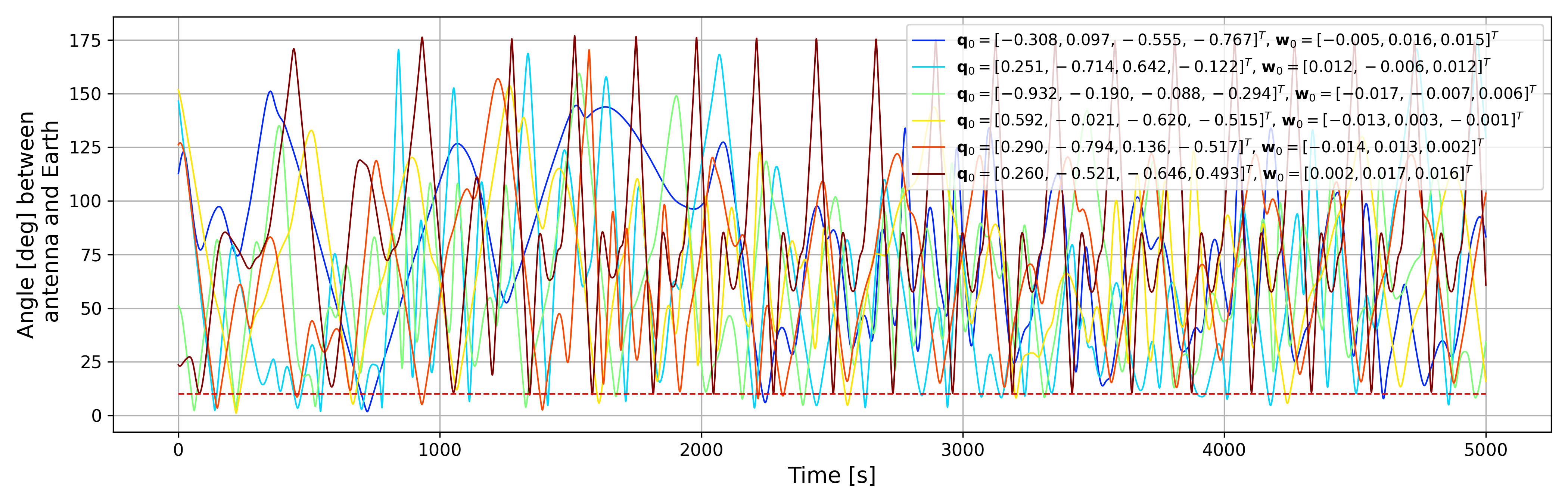}
    }
    \caption{The evaluation of the angle difference $\alpha$ in various initial conditions (Experiment: Two Torque Failures)}
    \label{fig:twofailure}
    
\end{figure}

We find that when $M_y$ and $M_z$ are failed, the RL can still reorient the antenna towards the Earth for all the initial states, as shown in Figure \ref{fig:twofailure-b}. The performance for the other two cases is not as good. In Figure \ref{fig:twofailure-a} and Figure \ref{fig:twofailure-c}, for most cases, the antenna can point to the Earth for a short duration every once in a while, and the angle differences are close to the $10^\circ$ threshold. 

For comparison, even though Hao et al.\cite{peng2023reorient} got similar results in Figure \ref{fig:twofailure(hao)-a} than ours, our results show a smoother angle change in the other two experiments (shown in Figure \ref{fig:twofailure-b} and Figure \ref{fig:twofailure-c}).

\subsubsection{Performance in Experiments with Gyroscope Errors}

In this section, we demonstrate the performance of the RL agent under gyroscope noises, gyroscope drift, and constant gyroscope error, which are unknown to the RL agent. Hao's paper does not evaluate the trained model under gyroscope-related perturbations. For all the cases in this subsection, the RL agent has run 200,0000 iterations to get the final trained model.

\begin{figure}[htb]
    \centering
    \subfigure[Gyroscope Noise Experiment \label{fig:gyroscopenoise+constant-noise}]{
        \includegraphics[width=0.7\textwidth]{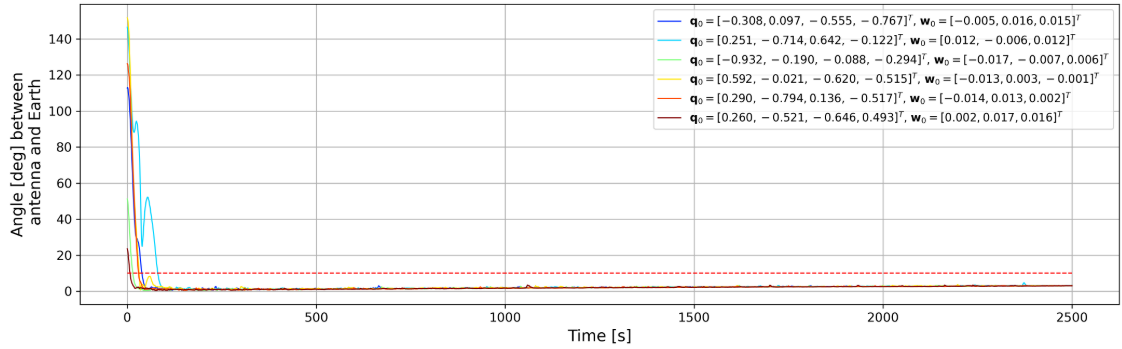}
    }
    \\ 
    \subfigure[Gyroscope Constant Error Experiment \label{fig:gyroscopenoise+constant-constant}]{
        \includegraphics[width=0.7\textwidth]{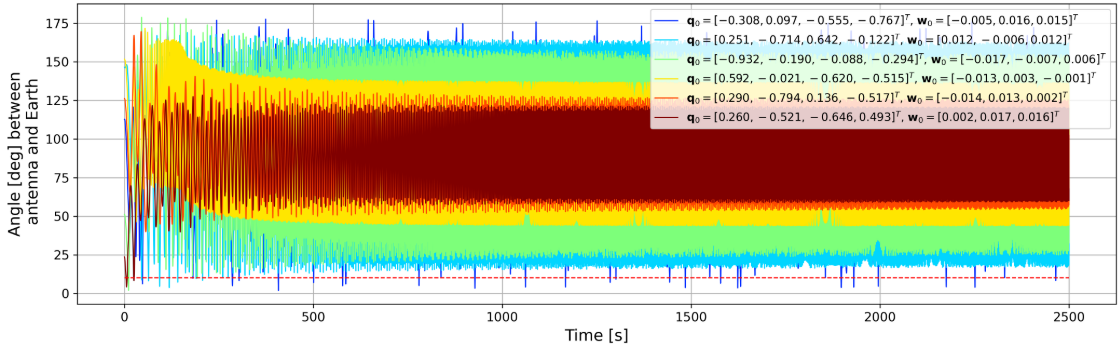}
    }
    \caption{The evaluation of the angle difference $\alpha$ in various initial conditions (Experiment: Gyroscope Noise Error and Gyroscope Constant Error)}
    \label{fig:gyroscopenoise+constant}
\end{figure}

As is shown in Figure \ref{fig:gyroscopenoise+constant}, the trained RL agent can handle random gyroscope noises but has bad performance while dealing with constant gyroscope errors that vary from each training episode. This is because these noises are randomly selected from a Gaussian distribution. When the number of samples from one distribution approaches infinity, the empirical distribution of the samples will converge to the true distribution. For the constant gyroscope error experiment, even though the gyroscope error remains the same during each episode, the insufficient number of total episodes results in insufficient gyroscope error types encountered by the agent to converge to a specific distribution. Therefore, the RL agent can't handle constant gyroscope errors after the same number of iterations as the gyroscope noise experiment.

\begin{figure}[htb]
    \centering
    \subfigure[Gyroscope Drift Experiment \label{fig:gyrodrift_0.01}]{
        \includegraphics[width=0.7\textwidth]{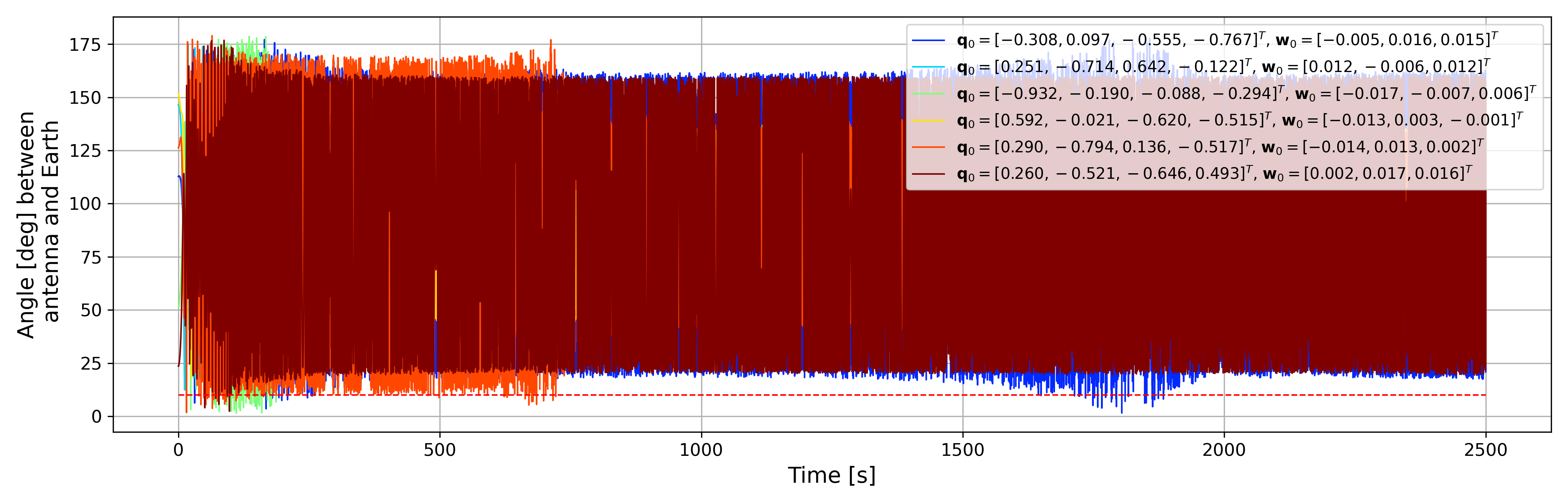}
    }
    \\ 
    \subfigure[Gyroscope Drift Experiment \label{fig:gyrodrift_0.001}]{
        \includegraphics[width=0.7\textwidth]{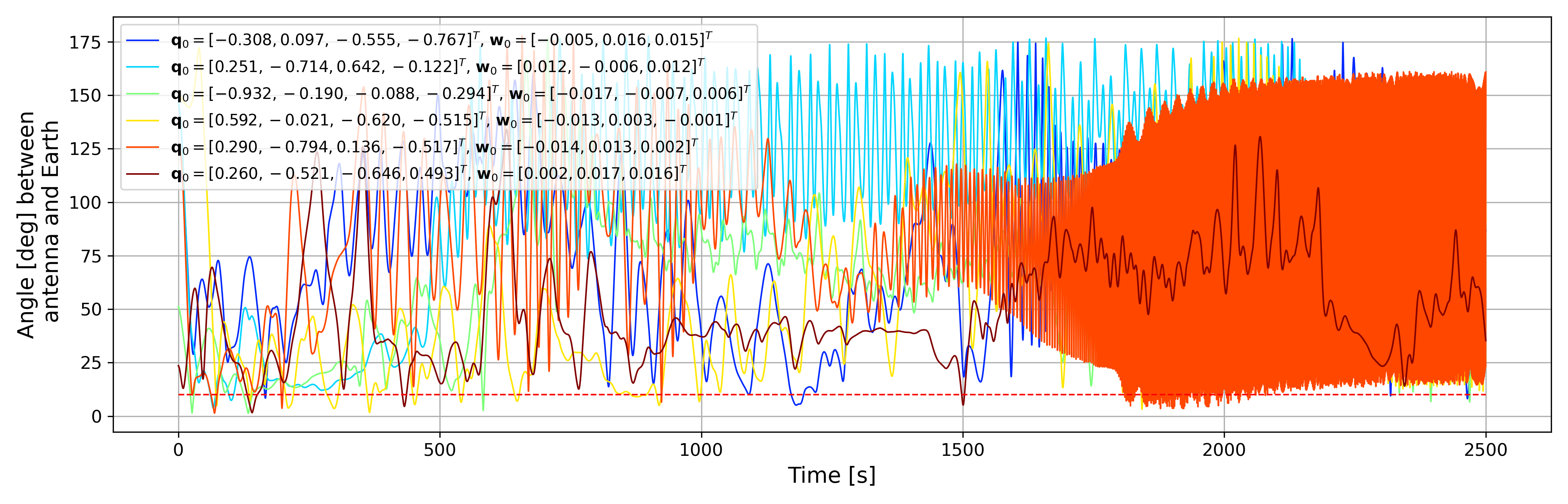}
    }
    \\ 
    \subfigure[Gyroscope Drift Experiment \label{fig:gyrodrift_0.0001}]{
        \includegraphics[width=0.7\textwidth]{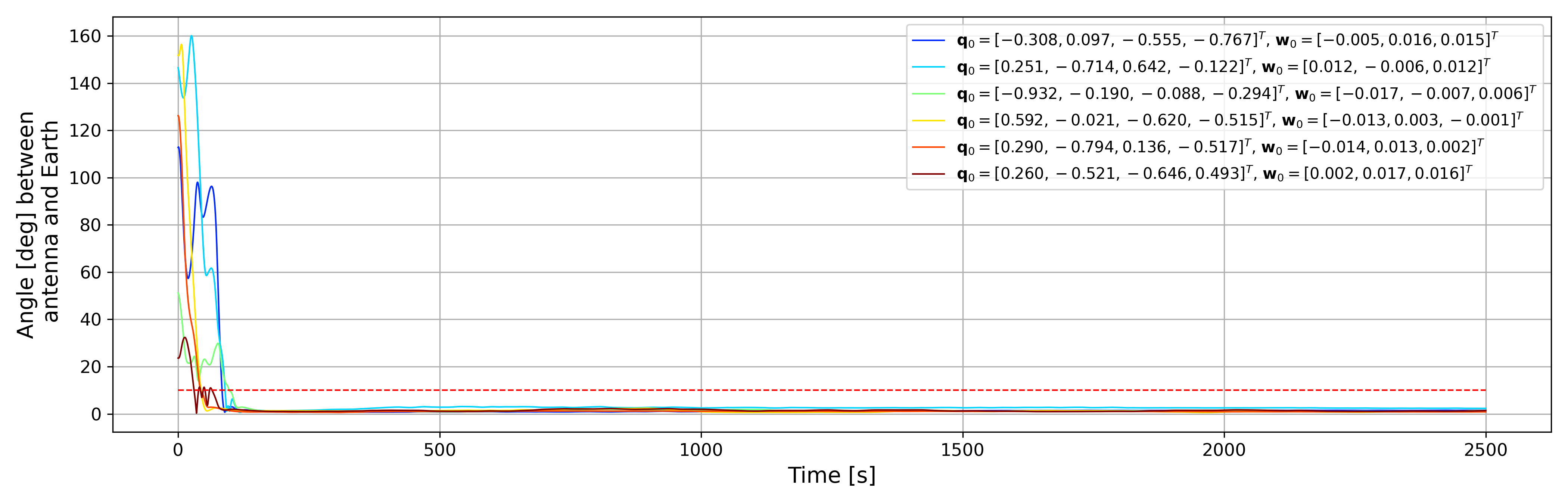}
    }
    \caption{The evaluation of the angle difference $\alpha$ in various initial conditions (Experiment: Gyroscope Drift Errors)}
    \label{fig:gyrodrift}
\end{figure}

For the gyroscope drift experiment, the performance of the trained RL agent depends on the magnitude of the $\sigma_b$ mentioned in the definition of the gyroscope drift perturbation. As we can see in Figure \ref{fig:gyrodrift}, the agent has a harder time adapting to the gyroscope drift perturbation than the previous gyroscope noise perturbation. Because it can deal with the gyroscope noise perturbation with $\sigma_\eta = 0.1^\circ/s$, but it doesn't work for the gyroscope drift perturbation until the $\sigma_b$ decreases to $0.0001^\circ/s$.

As we mentioned above, we have refined the reward function and used the SAC algorithm, which led to better results than Hao et al.'s work \cite{peng2023reorient}. The analysis attached in Appendix C discusses the ablation study, removing the change of the reward function or the SAC algorithm to verify their contribution to our trained SAC expert algorithm.

\subsection{GAIL Algorithm}

To verify the efficiency of the GAIL algorithm and minimize the differences other than the algorithm itself, the hyperparameters of the learner and the expert of the GAIL algorithm are identical. Following the index used in Table \ref{tab:PerExperiment}, we evaluate the performance of the learner in experiments 1, 2, 3, 4, 5, 6, 7, 8, 10, and 11. Experiments 9, 12, 13, and 14 are not included because, as previous experiments in the last section showed, the expert itself can't handle these perturbations. There is little point in conducting these evaluations.

\subsubsection{Performance of the Learner during Training}

We monitor the training process of the learner and record its mean reward value. Then, we compare it with what happens during the training process of the expert. This allows us to evaluate the learner's training efficiency.

\begin{figure}[htb]
    \centering
    \includegraphics[width=1\linewidth]{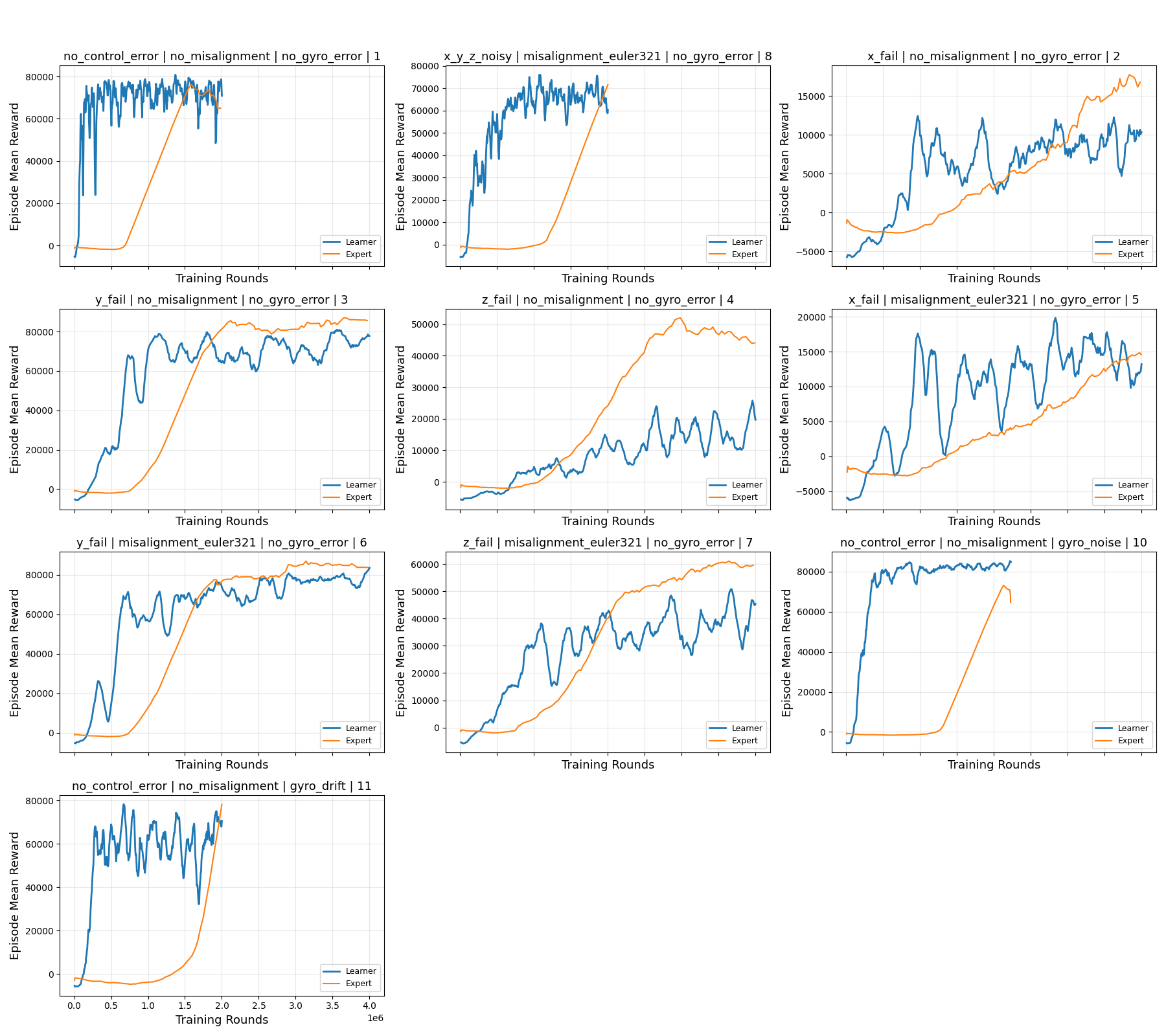}
    \caption{Mean reward value during the training process of the learner and the expert}
    \label{fig:EpisodeRewardExpertLearner}
\end{figure}

As we can see in Figure \ref{fig:EpisodeRewardExpertLearner}, the training efficiency of the learner is higher than that of the expert in all experiments, even though the learning efficiency of learners varies in different experiments. It can gain a higher mean reward in the early stage of the training than the expert. 

In terms of the highest reward value obtained throughout the training process, the learner can get a similar value as the expert in Experiments 1, 8, 6, and 11, even higher value in experiments 5 and 10. For other Experiments, even though the learner can't obtain a similar value as the expert, its learning efficiency is higher than the expert's.

We also conduct the ablation study for the GAIL algorithm. Results are shown in Appendix D.

\subsubsection{Performance of the Learner in Experiments}

After showing the comparison of the mean reward value, we conduct several evaluation experiments for the learner. For each case, we apply the six randomly generated initial states to the dedicated learner. The setting of these six initial states is identical to that in the experiment of the expert.

\begin{figure}[htb]
    \centering
    \includegraphics[width=0.7\linewidth]{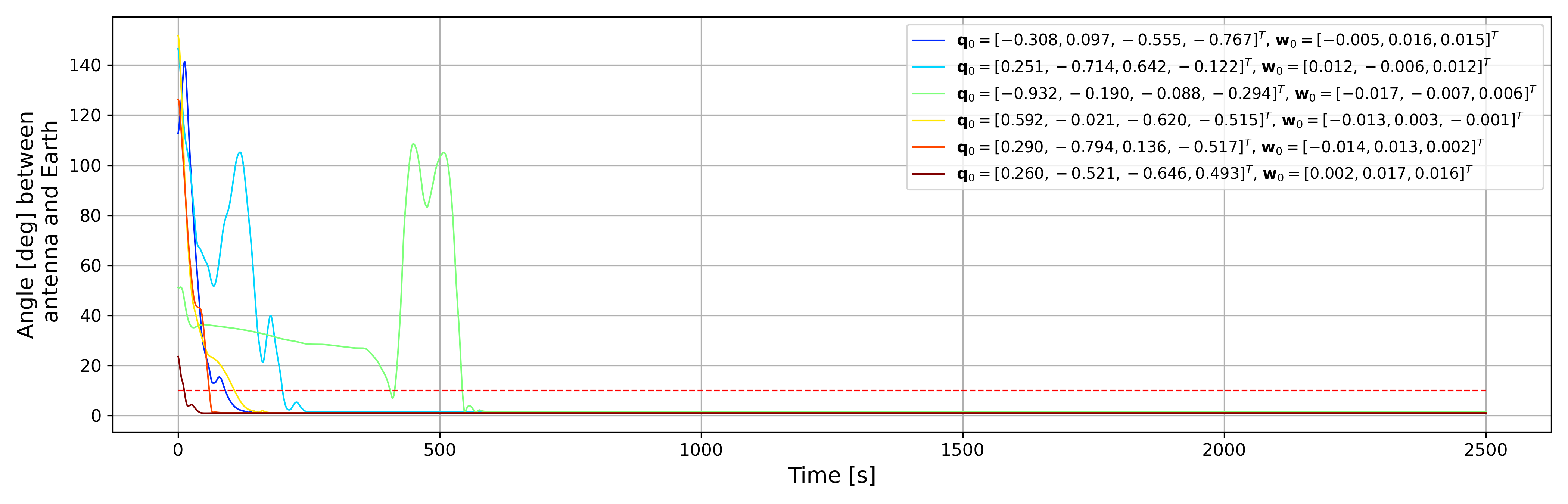}
    \caption{The evaluation of the angle difference $\alpha$ in various initial conditions (Experiment 1)}
    \label{fig:baseline(gail)}
\end{figure}

\begin{table}[htb]
    \centering
    \begin{tabular}{c|c|c|c|c}
        \hline 
         & RMS & Duty Cycle & Max ($^\circ$) & Min ($^\circ$) \\
        \hline 
        Learner & 13.7349 & 0.9331 & 116.3232 & 1.1510 \\
        Expert & 10.6214 & 0.9824 & 110.3312 & 0.8788 \\
        \hline 
    \end{tabular}
    \caption{Quantitative Evaluation for Experiment 1}
    \label{tab:QE1}
\end{table}

If we just observe Figure \ref{fig:baseline} and Figure \ref{fig:baseline(gail)}, the performance of the learner is as good as that of the expert. Even though the quantitative evaluation shows that the learner is slightly worse than the expert in the Table \ref{tab:QE1}, the learner has completed the goal we claimed in Experiment 1.

\begin{figure}[htb]
    \centering
    \includegraphics[width=0.7\linewidth]{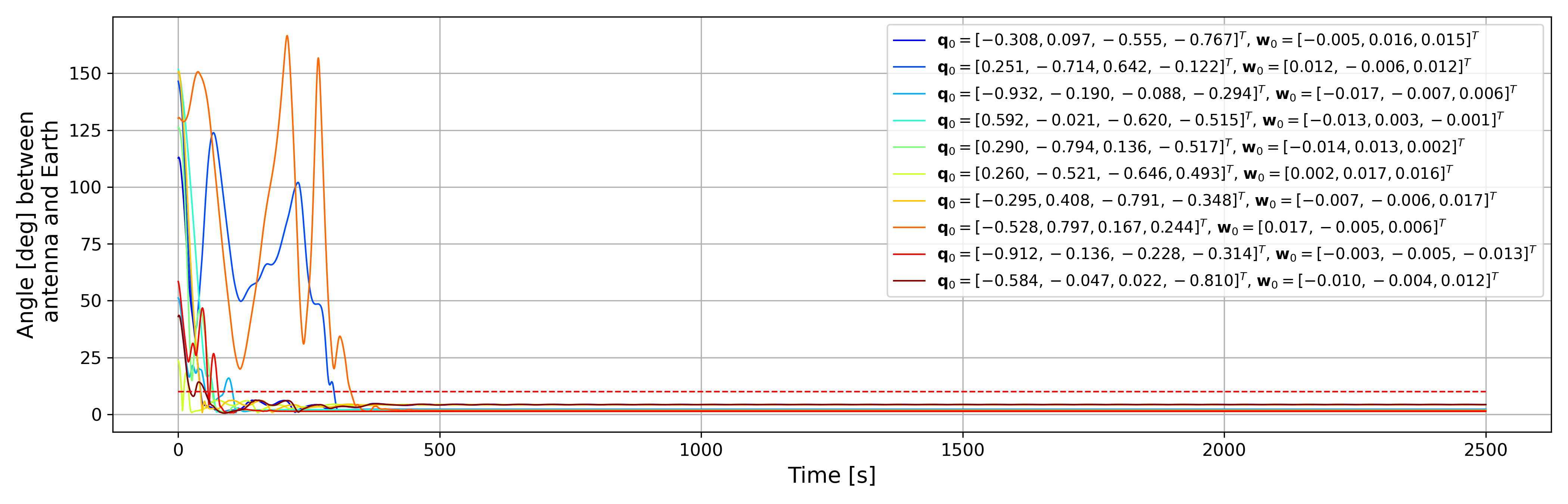}
    \caption{The evaluation of the angle difference $\alpha$ in various initial conditions (Experiment 8)}
    \label{fig:torquenoise+misalignment(gail)}
\end{figure}

\begin{table}[htb]
    \centering
    \begin{tabular}{c|c|c|c|c}
        \hline 
         & RMS & Duty Cycle & Max ($^\circ$) & Min ($^\circ$) \\
        \hline 
        Learner & 13.0336 & 0.9583 & 103.1712 & 0.8776 \\
        Expert & 9.7284 & 0.9838 & 103.1190 & 0.6873 \\
        \hline 
    \end{tabular}
    \caption{Quantitative Evaluation for Experiment 8}
    \label{tab:QE8}
\end{table}

For Experiment 8, as we can see in Figure \ref{fig:torquenoise+misalignment(gail)}, Figure \ref{fig:torquenoise+misalignment}, and Table \ref{tab:QE8}, a similar conclusion can also be made as we got in Experiment 1. The trained learner can handle the perturbation with the perturbation of the torque noise and the attitude misalignment.

\begin{figure}[htb]
    \centering
    \subfigure[$M_x$ Failure Experiment \label{fig:singlefailure(gail)-a}]{
        \includegraphics[width=0.7\textwidth]{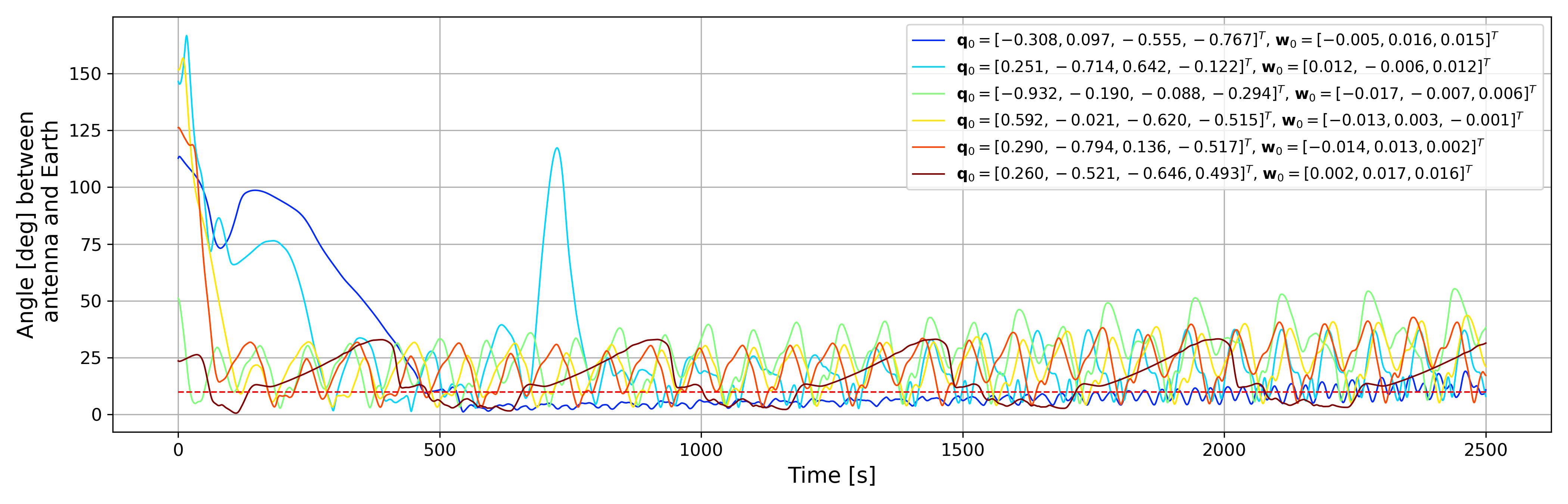}
    }
    \\ 
    \subfigure[$M_y$ Failure Experiment \label{fig:singlefailure(gail)-b}]{
        \includegraphics[width=0.7\textwidth]{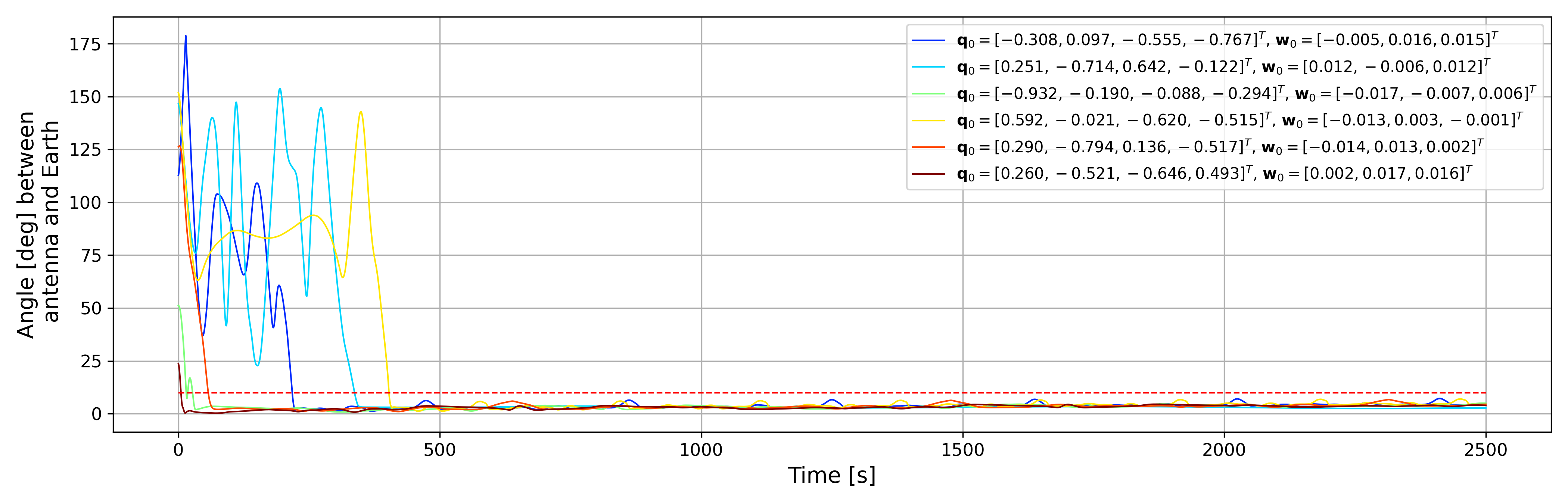}
    }
    \\ 
    \subfigure[$M_z$ Failure Experiment \label{fig:singlefailure(gail)-c}]{
        \includegraphics[width=0.7\textwidth]{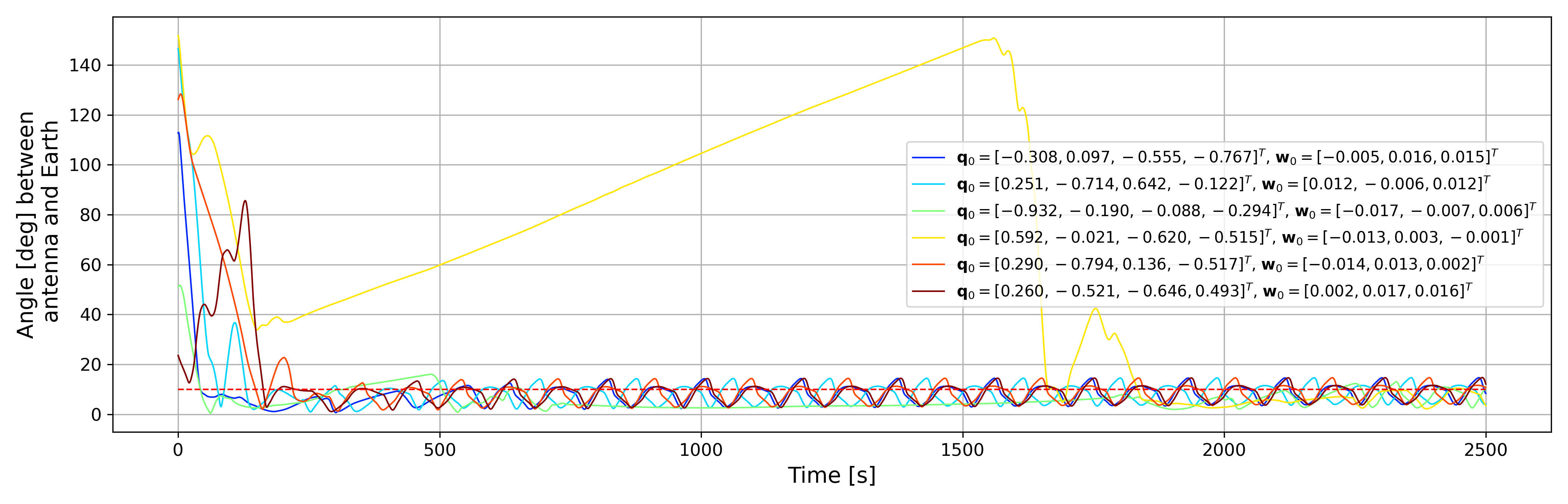}
    }
    \caption{The evaluation of the angle difference $\alpha$ in various initial conditions (Experiment 2, 3, and 4)}
    \label{fig:singlefailure(gail)}
    
\end{figure}

\begin{table}[htb]
    \centering
    \begin{tabular}{c|c|c|c|c}
        \hline 
         & RMS & Duty Cycle & Max ($^\circ$) & Min ($^\circ$) \\
        \hline 
        Learner in Experiment 2 & 29.2109 & 0.2565 & 108.6495 & 2.0804 \\
        Expert in Experiment 2 & 30.6388 & 0.3470 & 109.8171 & 1.7893 \\
        Learner in Experiment 3 & 19.8102 & 0.9304 & 114.2845 & 1.2022 \\
        Expert in Experiment 3 & 6.6081 & 0.9896 & 81.3018 & 0.4272 \\
        Learner in Experiment 4 & 25.8185 & 0.6026 & 112.8235 & 1.1303 \\
        Expert in Experiment 4 & 14.2068 & 0.8670 & 102.2431 & 1.0300 \\
        \hline 
    \end{tabular}
    \caption{Quantitative Evaluation for Experiment 2, 3, and 4}
    \label{tab:QE234}
\end{table}

As we can see in Figure \ref{fig:singlefailure(gail)}, Figure \ref{fig:singlefailure}, and Table \ref{tab:QE234}, although the learner performs worse than the expert in quantitative evaluation, it can achieve similar or even better results as the expert in single torque failure experiments. As Experiment 2 showed in Figure \ref{fig:singlefailure(gail)-a}, the learner shows less RMS error and less Maximum error than the expert shown in Figure \ref{fig:singlefailure-a}. Even though there are some oscillations in Figure \ref{fig:singlefailure(gail)-b} and Figure \ref{fig:singlefailure(gail)-b}, the learner's performance is not much different from the expert's.

\begin{figure}[htb]
    \centering
    \subfigure[$M_x$ Failure and Attitude Misalignment Experiment \label{fig:singlefailure+misalignment(gail)-a}]{
        \includegraphics[width=0.7\textwidth]{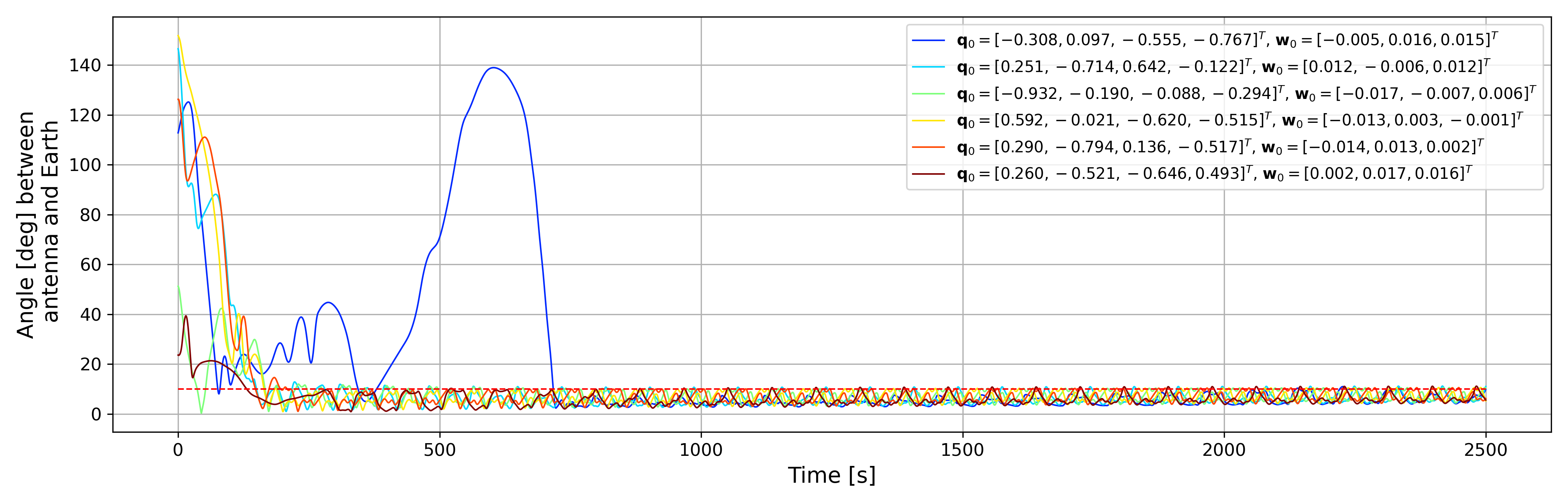}
    }
    \\ 
    \subfigure[$M_y$ Failure and Attitude Misalignment Experiment \label{fig:singlefailure+misalignment(gail)-b}]{
        \includegraphics[width=0.7\textwidth]{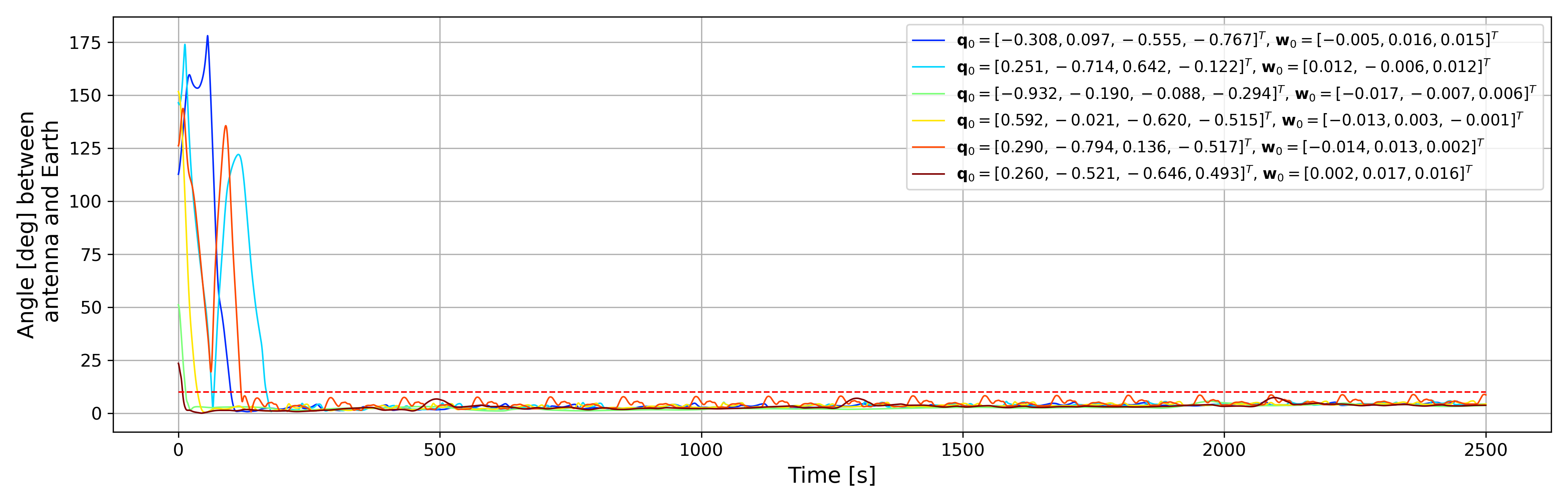}
    }
    \\ 
    \subfigure[$M_z$ Failure and Attitude Misalignment Experiment \label{fig:singlefailure+misalignment(gail)-c}]{
        \includegraphics[width=0.7\textwidth]{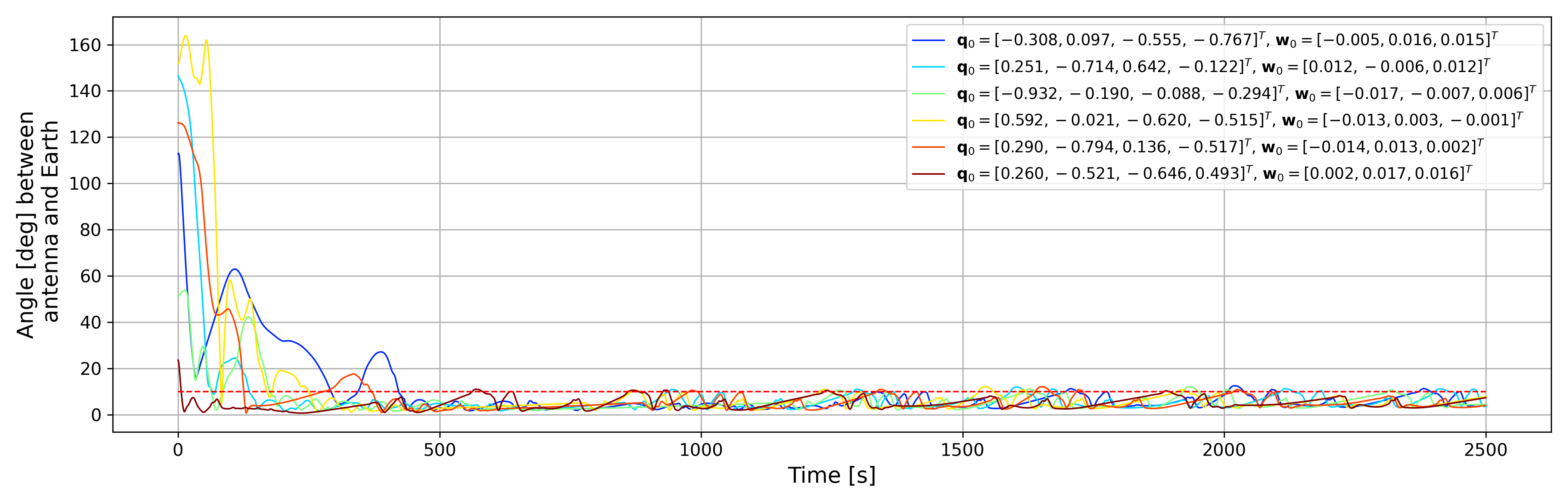}
    }
    \caption{The evaluation of the angle difference $\alpha$ in various initial conditions (Experiment 5, 6, and 7)}
    \label{fig:singlefailure+misalignment(gail)}
\end{figure}

\begin{table}[htb]
    \centering
    \begin{tabular}{c|c|c|c|c}
        \hline 
         & RMS & Duty Cycle & Max ($^\circ$) & Min ($^\circ$) \\
        \hline 
        Learner in Experiment 5 & 20.0112 & 0.8474 & 108.9920 & 1.0352 \\
        Expert in Experiment 5 & 24.2443 & 0.4133 & 104.5597 & 1.5482 \\
        Learner in Experiment 6 & 15.0766 & 0.9704 & 120.4200 & 0.6906 \\
        Expert in Experiment 6 & 8.5416 & 0.9890 & 102.0269 & 0.5508 \\
        Learner in Experiment 7 & 15.86 & 0.8862 & 104.4968 & 1.2182 \\
        Expert in Experiment 7 & 18.6904 & 0.9329 & 110.3715 & 0.9411 \\
        \hline 
    \end{tabular}
    \caption{Quantitative Evaluation for Experiment 5, 6, and 7}
    \label{tab:QE567}
\end{table}

As we can see in Figure \ref{fig:singlefailure+misalignment(gail)}, Figure \ref{fig:singlefailure+misalignment}, and Table \ref{tab:QE567}, the performance of the learner and the expert is almost the same. Although the single torque failure with the attitude misalignment perturbation is more difficult for the RL agent to handle than the perturbation of the single torque failure itself, the GAIL algorithm is essentially a data-driven method. Its performance is more directly related to the quality of the trajectory dataset. Therefore, as long as the expert algorithm responsible for generating the dataset performs well, the GAIL algorithm is less susceptible to more difficult perturbations.

\begin{figure}[htb]
    \centering
    \subfigure[Gyroscope Noise Experiment \label{fig:gyroscopenoise(gail)}]{
        \includegraphics[width=0.7\textwidth]{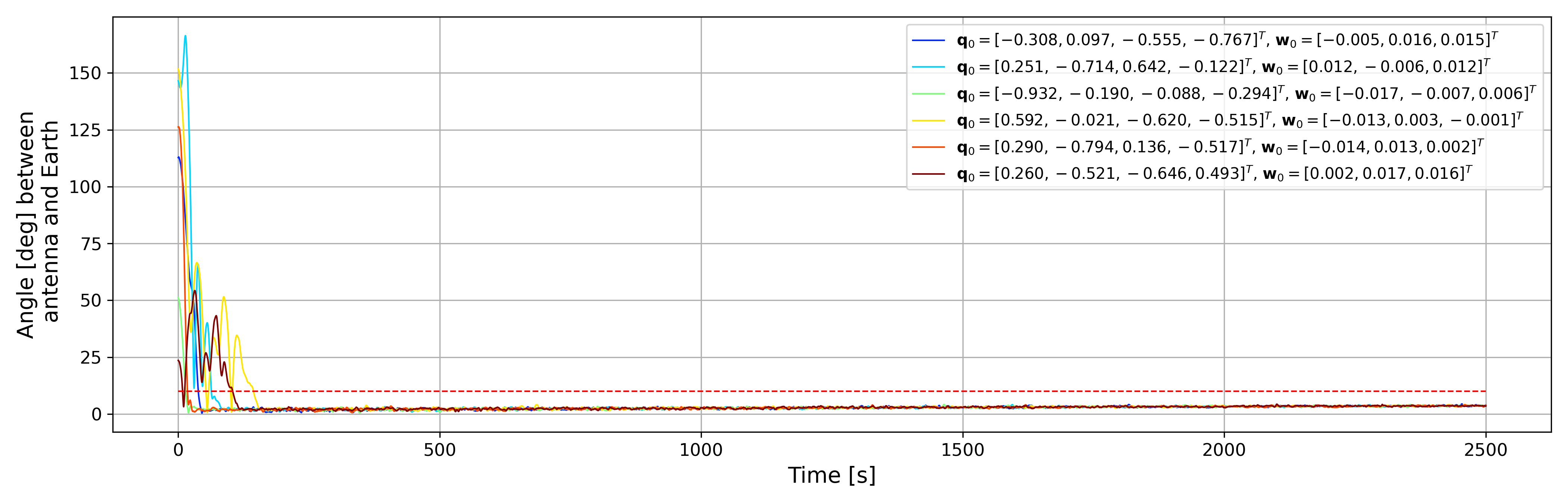}
    }
    \\ 
    \subfigure[Gyroscope Drift Experiment \label{fig:gyroscopenoisedrift(gail)}]{
        \includegraphics[width=0.7\textwidth]{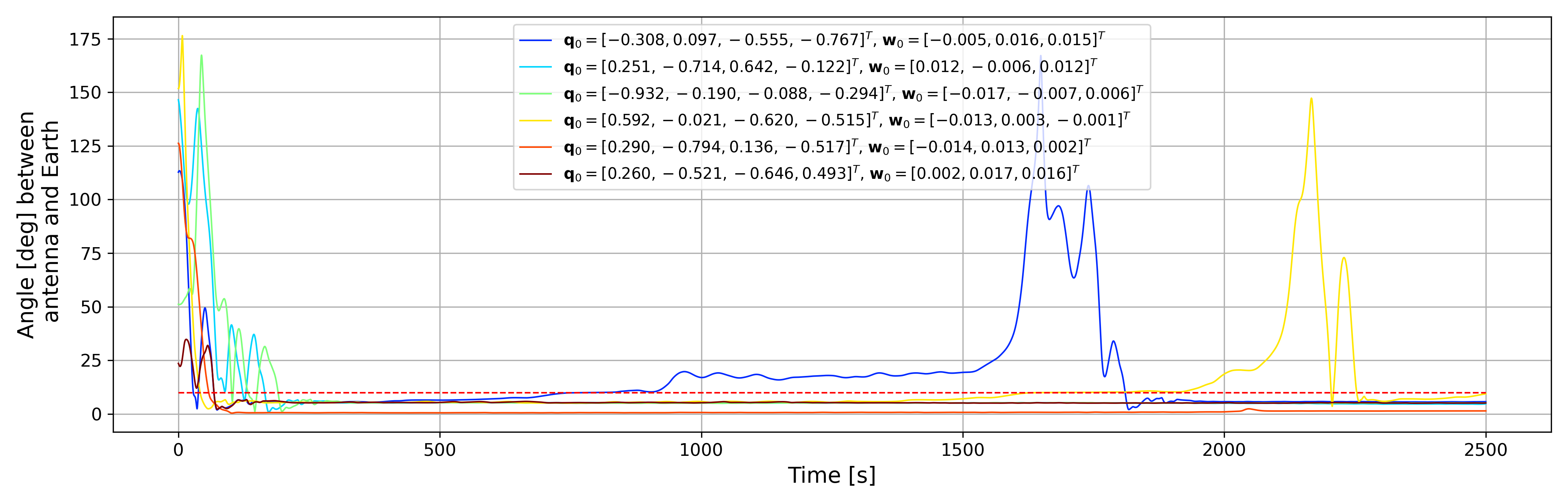}
    }
    \caption{The evaluation of the angle difference $\alpha$ in various initial conditions (Experiment 10 and 11)}
    \label{fig:GyroscopeNoiseDrift}
\end{figure}

\begin{table}[htb]
    \centering
    \begin{tabular}{c|c|c|c|c}
        \hline 
         & RMS & Duty Cycle & Max ($^\circ$) & Min ($^\circ$) \\
        \hline 
        Learner in Experiment 10 & 9.6879 & 0.9754 & 110.4877 & 0.7383 \\
        Expert in Experiment 10 & 8.7578 & 0.9861 & 104.2707 & 0.3524 \\
        Learner in Experiment 11 & 18.2910 & 0.8546 & 136.3973 & 1.5027 \\
        Expert in Experiment 11 & 10.9631 & 0.9732 & 106.5313 & 0.6092 \\
        \hline 
    \end{tabular}
    \caption{Quantitative Evaluation for Experiment 10 and 11}
    \label{tab:QE1011}
\end{table}

As we can see in Figure \ref{fig:GyroscopeNoiseDrift}, Figure \ref{fig:gyrodrift}, and Table \ref{tab:QE1011}, the learner can easily achieve the same effect as an expert in the gyroscope noise perturbation experiments. For the gyroscope drift experiment, there is a little abnormal swinging that happens in the satellite controlled by the learner in the second half of the evaluation. However, overall, it successfully imitates the expert's behavior.


\section{Conclusion}

This paper addresses the challenge of satellite attitude control by proposing an innovative approach integrating reinforcement learning (RL) and Generative Adversarial Imitation Learning (GAIL). Through the design and adjustment of the reward function, the RL agent demonstrates enhanced performance, achieving superior stability and efficiency across diverse perturbation experiments compared to an earlier study. Additionally, the incorporation of GAIL further allows the agent to effectively generalize from expert demonstrations, dramatically decreasing the sample complexity and computational demand associated with traditional RL training approaches.

\section{Limitation and Future Work}

The proposed method opens new avenues in the domain of spacecraft attitude management, presenting a pathway toward more autonomous, efficient, and resilient control systems. The current hyperparameters of the GAIL algorithm vary in different experiments. This may be because the distribution of expert datasets in different experiments is different. It is necessary to fine-tune the hyperparameters so that GAIL can have a certain degree of adaptability in different experiments. We will try to develop a model based on the standard GAIL algorithm that can perform significantly better than the expert and adapt to more complicated gyroscope-related perturbations.

Future research directions include further optimization of the adversarial attack strategy, exploration of real-world deployment scenarios, and expansion to more complex multi-satellite coordination tasks. The successful application of these advanced learning techniques signifies a promising step forward in astrodynamics and space-flight mechanics, contributing significantly to the development of intelligent satellite control systems.

\section{Acknowledgment}

The research has been supported by the Air Force Office of Scientific Research (AFOSR) grant number FA9550-22-1-0364.

\appendix
\section{Appendix A: Hyperparameters of Our Method}
\label{appendix:Hyperparameter}

\subsection*{Expert in GAIL}

Table \ref{tab:HPExpert} shows the hyperparameters used in our trained SAC expert algorithm.

\begin{table}[htb]
    \fontsize{10}{10}\selectfont
    \caption{List of Hyperparameters used in SAC Expert Algorithm}
    \label{tab:HPExpert}
        \centering 
    \begin{tabular}{c | r | r } 
        \hline 
        Hyperparameter & Value & Description\\
        \hline 
        learning rate & $0.001 \xrightarrow{} 0.0001$ & Rate at which the neural network learns \\
        batch size & 64 & Number of samples per gradient update \\
        buffer size & 1000000 & Size of experience replay buffer \\
        learning starts & 1000 & Initial steps before starting training \\
        train frequency & 1 & Frequency of performing training updates \\
        gradient steps & 1 & Number of gradient updates per training step \\
        policy & MLP policy & Specifies the type of neural network used as the policy \\
        \hline
    \end{tabular}
\end{table}

\section{Appendix B: Results from the Previous Work}
\label{appendix:PreviousResults}

This appendix presents supplementary figures from Hao et al.'s work\cite{peng2023reorient} that are relevant to our current study. These figures provide additional context and demonstrate prior achievements in the development of related systems and methodologies.

\begin{figure}[htb]
    \centering
    \subfigure[$M_x$ Failure Experiment \label{fig:singlefailure(hao)-a}]{
        \includegraphics[width=0.7\textwidth]{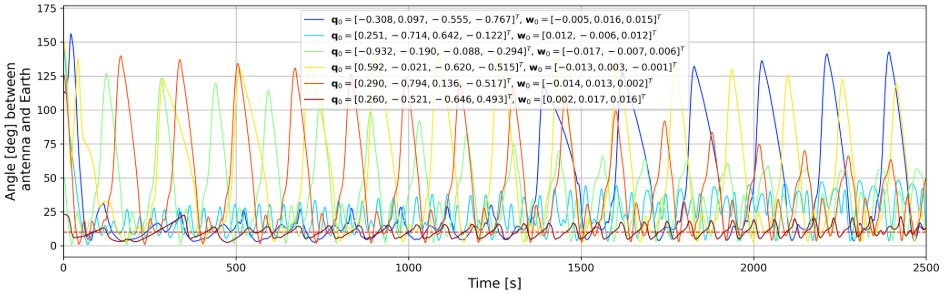}
    }
    \\ 
    \subfigure[$M_y$ Failure Experiment \label{fig:singlefailure(hao)-b}]{
        \includegraphics[width=0.7\textwidth]{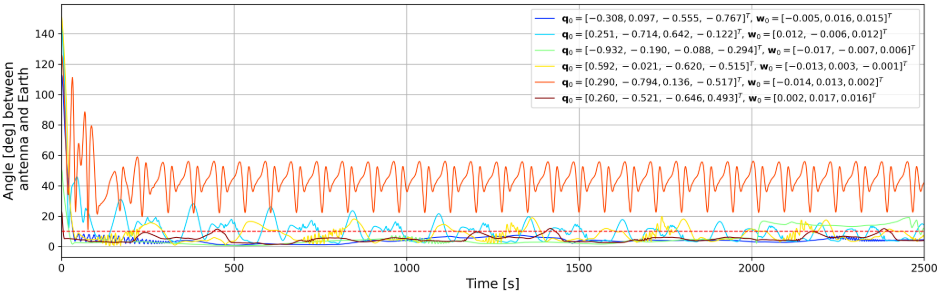}
    }
    \\ 
    \subfigure[$M_z$ Failure Experiment \label{fig:singlefailure(hao)-c}]{
        \includegraphics[width=0.7\textwidth]{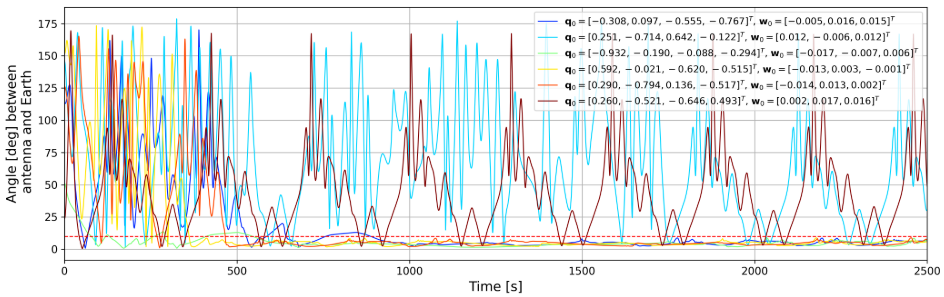}
    }
    \caption{The evaluation of the angle difference $\alpha$ in various initial conditions (Experiment: Single Torque Failure in Hao's \cite{peng2023reorient} work)}
    \label{fig:singlefailure(hao)}
\end{figure}

\begin{figure}[htb]
    \centering
    \includegraphics[width=0.7\linewidth]{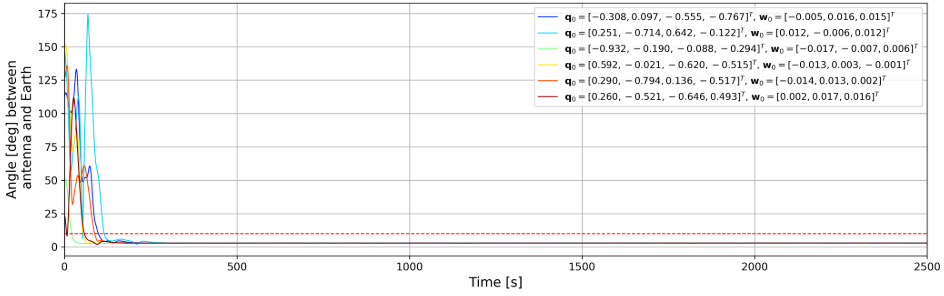}
    \caption{The evaluation of the angle difference $\alpha$ in various initial conditions (Experiment: Torque Noises and Attitude Misalignment in Hao's \cite{peng2023reorient} work)}
    \label{fig:torquenoise+misalignment(hao)}
\end{figure}

\begin{figure}[htb]
    \centering
    \subfigure[$M_x$ Failure and Attitude Misalignment Experiment \label{fig:singlefailure+misalignment(hao)-a}]{
        \includegraphics[width=0.7\textwidth]{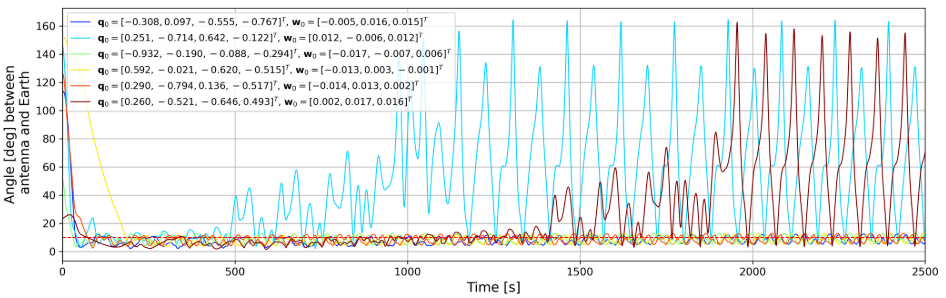}
    }
    \\ 
    \subfigure[$M_y$ Failure and Attitude Misalignment Experiment \label{fig:singlefailure+misalignment(hao)-b}]{
        \includegraphics[width=0.7\textwidth]{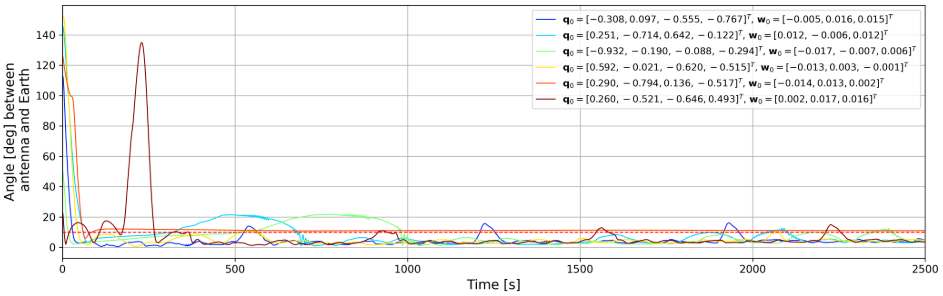}
    }
    \\ 
    \subfigure[$M_z$ Failure and Attitude Misalignment Experiment \label{fig:singlefailure+misalignment(hao)-c}]{
        \includegraphics[width=0.7\textwidth]{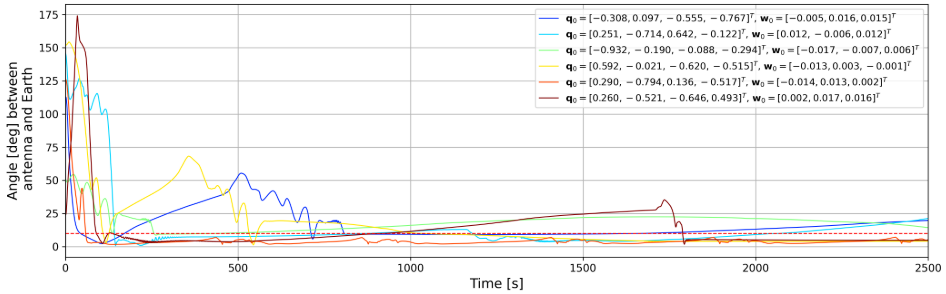}
    }
    \caption{The evaluation of the angle difference $\alpha$ in various initial conditions (Experiment: Single Torque Failure and Constant Attitude Misalignment in Hao's \cite{peng2023reorient} work)}
    \label{fig:singlefailure+misalignment(hao)}
\end{figure}

\begin{figure}[htb]
    \centering
    \subfigure[$M_x$ and $M_y$ Failure Experiment \label{fig:twofailure(hao)-a}]{
        \includegraphics[width=0.7\textwidth]{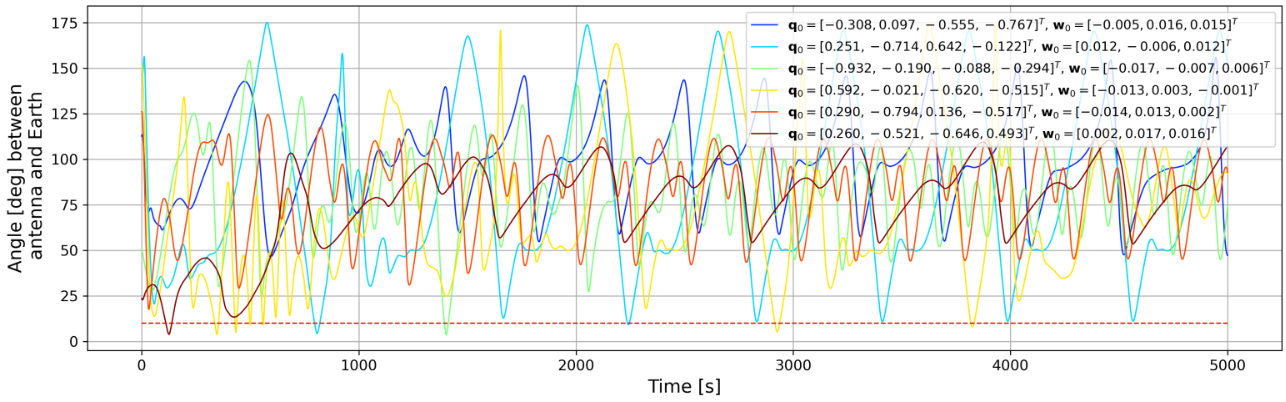}
    }
    \\ 
    \subfigure[$M_y$ and $M_z$ Failure Experiment \label{fig:twofailure(hao)-b}]{
        \includegraphics[width=0.7\textwidth]{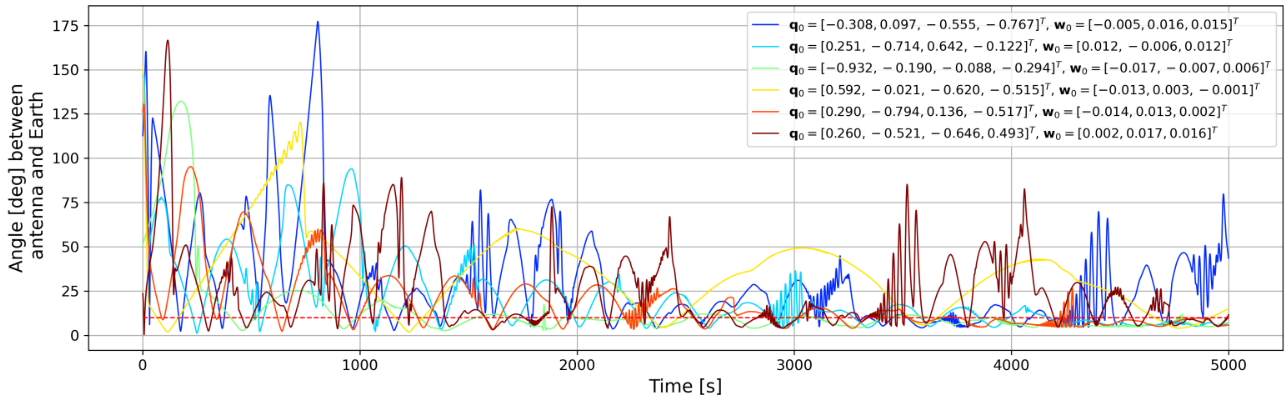}
    }
    \\ 
    \subfigure[$M_x$ and $M_z$ Failure Experiment \label{fig:twofailure(hao)-c}]{
        \includegraphics[width=0.7\textwidth]{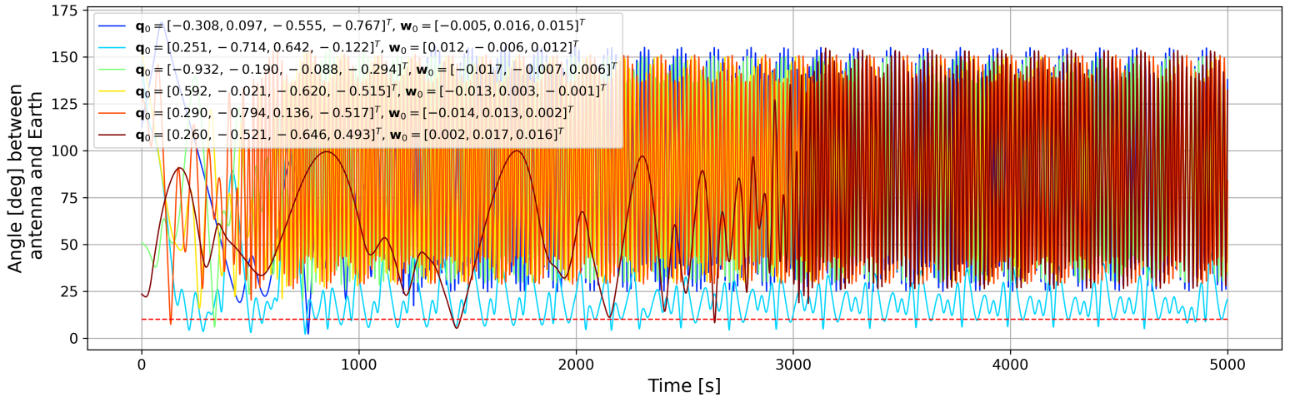}
    }
    \caption{The evaluation of the angle difference $\alpha$ in various initial conditions (Experiment: Two Torque Failures in Hao's \cite{peng2023reorient} work)}
    \label{fig:twofailure(hao)}
    
\end{figure}

\section{Appendix C: Additional Ablation Study for the SAC Expert Algorithm}
\label{appendix:PreviousResults}

Considering there are two changes in our trained SAC expert algorithm compared with Hao et al.'s work \cite{peng2023reorient} , the ablation study is divided into two parts: removing the SAC algorithm and removing the refined reward function.

\subsection{Removing the SAC Algorithm}

\begin{figure}[htb]
    \centering
    \subfigure[$M_x$ Failure Experiment \label{fig:singlefailure(DDPG)-x}]{
        \includegraphics[width=0.7\textwidth]{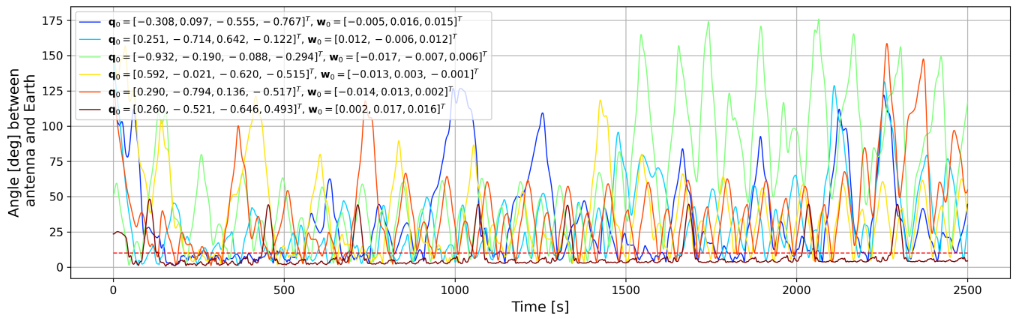}
    }
    \\ 
    \subfigure[$M_y$ Failure Experiment \label{fig:singlefailure(DDPG)-y}]{
        \includegraphics[width=0.7\textwidth]{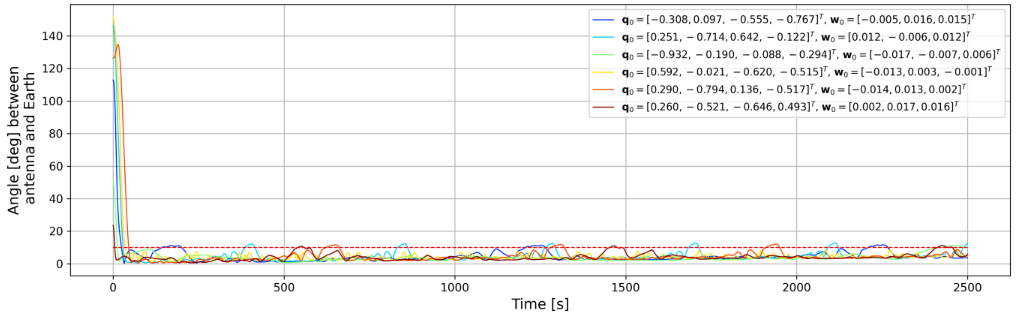}
    }
    \\ 
    \subfigure[$M_z$ Failure Experiment \label{fig:singlefailure(DDPG)-z}]{
        \includegraphics[width=0.7\textwidth]{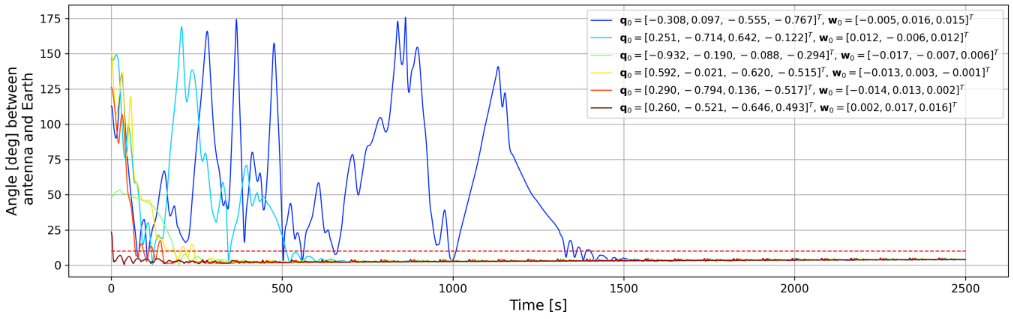}
    }
    \caption{The evaluation of the angle difference $\alpha$ in various initial conditions (Experiment: Single Torque Failure with the DDPG Algorithm and the Refined Reward Function)}
    \label{fig:singlefailure(DDPG)}
\end{figure}

\begin{figure}[htb]
    \centering
    \includegraphics[width=0.7\linewidth]{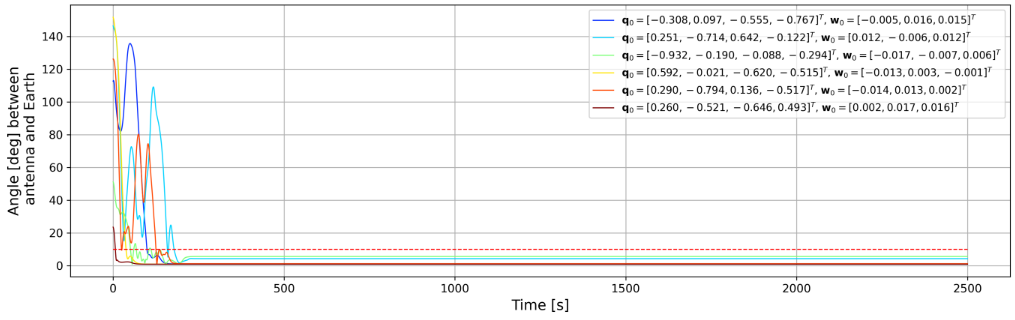}
    \caption{The evaluation of the angle difference $\alpha$ in various initial conditions (Experiment: Torque Noises and Attitude Misalignment with the DDPG Algorithm and the Refined Reward Function)}
    \label{fig:torquenoise+misalignment(DDPG)}
\end{figure}

\begin{figure}[htb]
    \centering
    \subfigure[$M_x$ Failure and Attitude Misalignment Experiment \label{fig:singlefailure+misalignment(DDPG)-x}]{
        \includegraphics[width=0.7\textwidth]{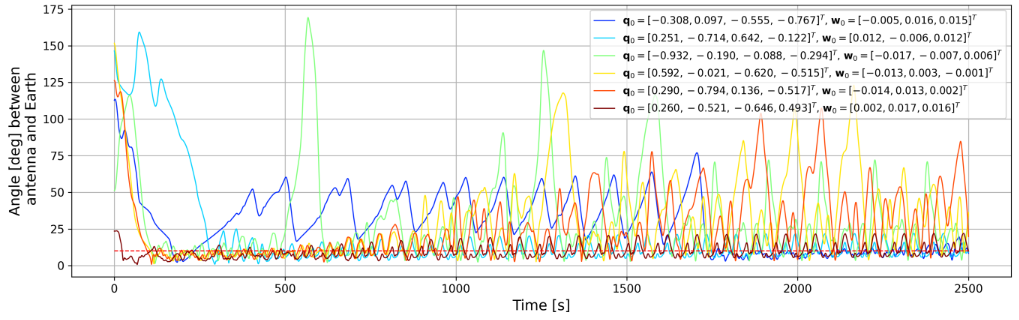}
    }
    \\ 
    \subfigure[$M_y$ Failure and Attitude Misalignment Experiment \label{fig:singlefailure+misalignment(DDPG)-y}]{
        \includegraphics[width=0.7\textwidth]{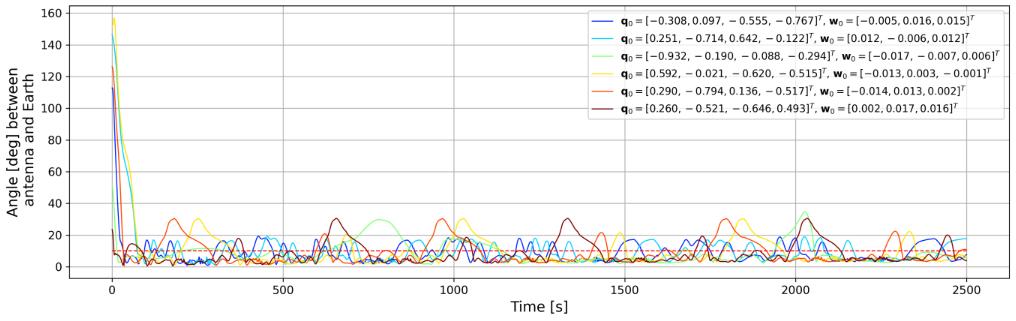}
    }
    \\ 
    \subfigure[$M_z$ Failure and Attitude Misalignment Experiment \label{fig:singlefailure+misalignment(DDPG)-z}]{
        \includegraphics[width=0.7\textwidth]{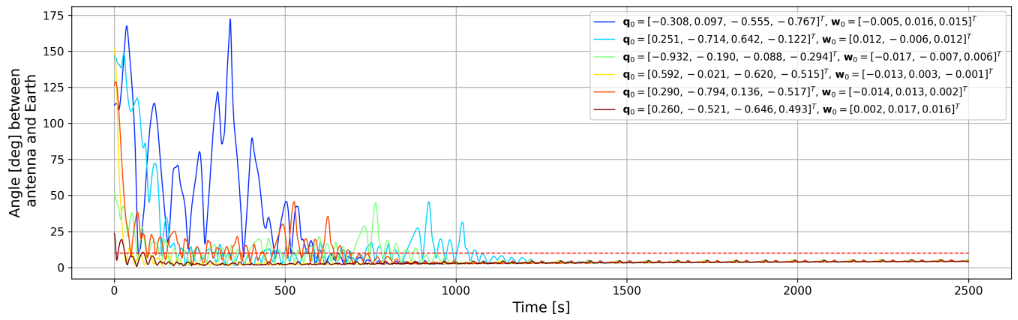}
    }
    \caption{The evaluation of the angle difference $\alpha$ in various initial conditions (Experiment: Single Torque Failure and Constant Attitude Misalignment with the DDPG Algorithm and the Refined Reward Function)}
    \label{fig:singlefailure+misalignment(DDPG)}
\end{figure}

\begin{figure}[htb]
    \centering
    \subfigure[Gyroscope Noise Experiment \label{fig:gyroscopenoise+constant(DDPG)-noise}]{
        \includegraphics[width=0.7\textwidth]{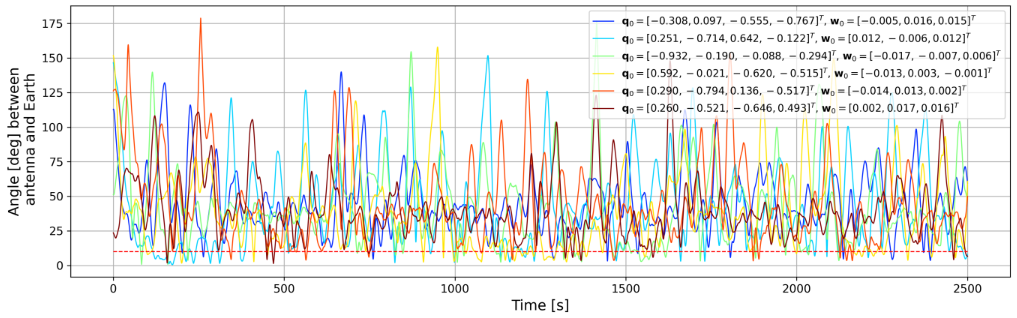}
    }
    \\ 
    \subfigure[Gyroscope Constant Error Experiment \label{fig:gyroscopenoise+constant(DDPG)-constant}]{
        \includegraphics[width=0.7\textwidth]{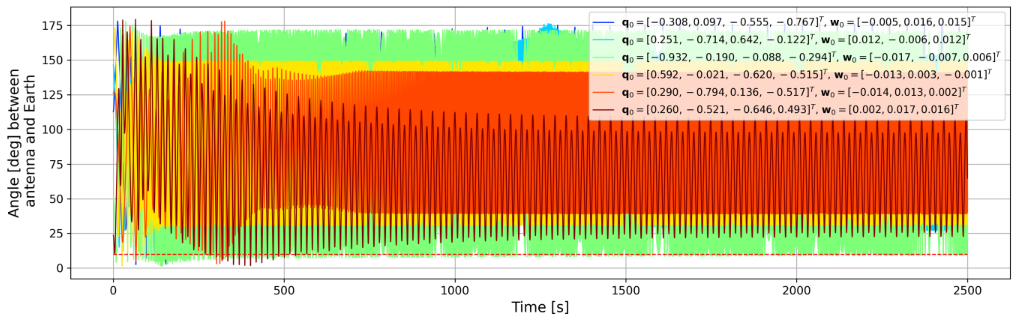}
    }
    \caption{The Evaluation of the Angular Difference $\alpha$ in Various Initial Conditions (Experiment: Gyroscope Errors with the DDPG Algorithm and the Refined Reward Function)}
    \label{fig:gyroscopenoise+constant(DDPG)}
\end{figure}

Here are the results (Figure \ref{fig:singlefailure(DDPG)}, Figure \ref{fig:torquenoise+misalignment(DDPG)}, Figure \ref{fig:singlefailure+misalignment(DDPG)}, and Figure \ref{fig:gyroscopenoise+constant(DDPG)}) with the DDPG algorithm and the refined reward function. 

Compared with using the SAC algorithm (Figure \ref{fig:singlefailure}), the results using the DDPG algorithm (Figure \ref{fig:singlefailure(DDPG)}) show that during the first half of the stage, the agent can't maintain the angle difference between the antenna and the target direction within the threshold for most of the cases in the $M_x$ Failure experiment. While there is an outlier that happens in Figure \ref{fig:singlefailure-c}, the agent achieves similar performance in $M_y$ and $M_z$ Failure Experiments.

As we can see in Figure \ref{fig:torquenoise+misalignment} and Figure \ref{fig:torquenoise+misalignment(DDPG)}, both the agents trained by the DDPG algorithm and the SAC algorithm have similarly successful results.

As is shown in Figure \ref{fig:singlefailure+misalignment(DDPG)}, compared with the agent trained by the SAC algorithm, the agent trained by the DDPG algorithm is able to achieve a similar percentage of the time that the angular error is within the threshold in all three experiments with both the torque errors and the attitude misalignment. However, the angular errors of many initial states show fluctuations with high frequency, especially for those in $M_z$ $M_y$ Failure and Attitude Misalignment experiment. This is because the agent trained by the DDPG algorithm only moves the antenna to the target position as much as possible, but cannot guarantee that the satellite will not rotate rapidly around the target position afterward.

Using the DDPG algorithm, the agent can't keep the antenna of the satellite within the threshold for the gyroscope noise experiment shown in Figure \ref{fig:gyroscopenoise+constant(DDPG)-noise}. The performance of the agent is significantly worse than that of the agent trained by the SAC algorithm (shown in Figure \ref{fig:gyroscopenoise+constant-noise}). For the gyroscope constant error experiment, both agents trained by the SAC and the DDPG algorithm can't control the satellite.

\subsection{Removing the Refined Reward Function}

\begin{figure}[htb]
    \centering
    \includegraphics[width=0.7\linewidth]{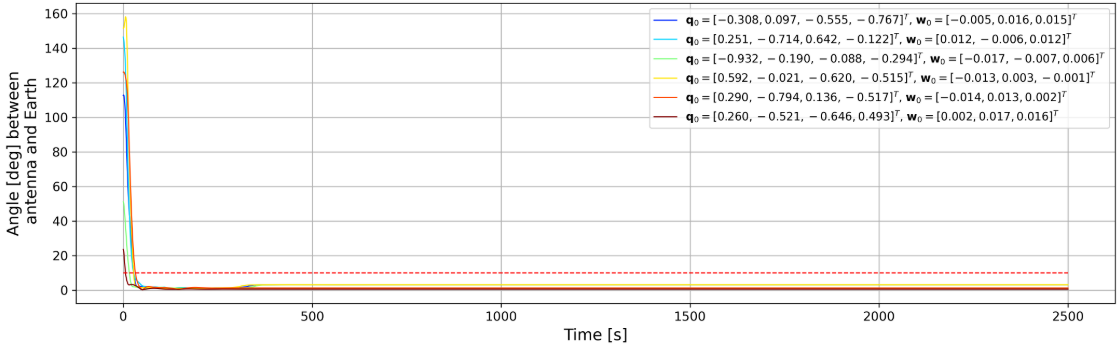}
    \caption{The evaluation of the angle difference $\alpha$ in various initial conditions (Experiment: Torque Noises and Attitude Misalignment with Previous Reward Function and SAC Algorithm)}
    \label{fig:baseline(p)}
\end{figure}

\begin{figure}[htb]
    \centering
    \subfigure[$M_x$ Failure Experiment \label{fig:singlefailure(p)-x}]{
        \includegraphics[width=0.7\textwidth]{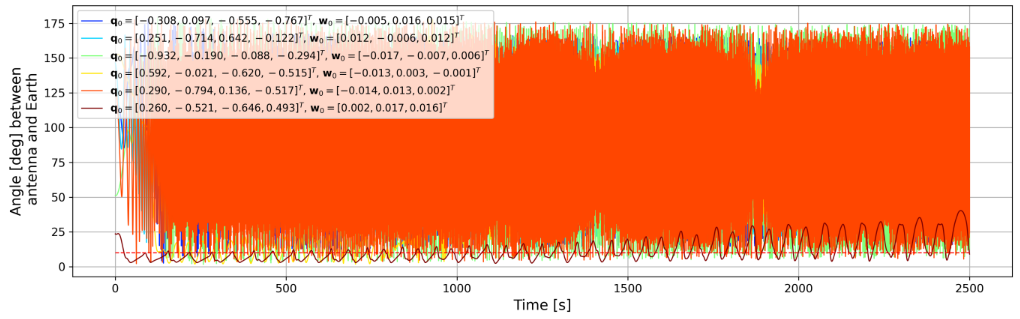}
    }
    \\ 
    \subfigure[$M_y$ Failure Experiment \label{fig:singlefailure(p)-y}]{
        \includegraphics[width=0.7\textwidth]{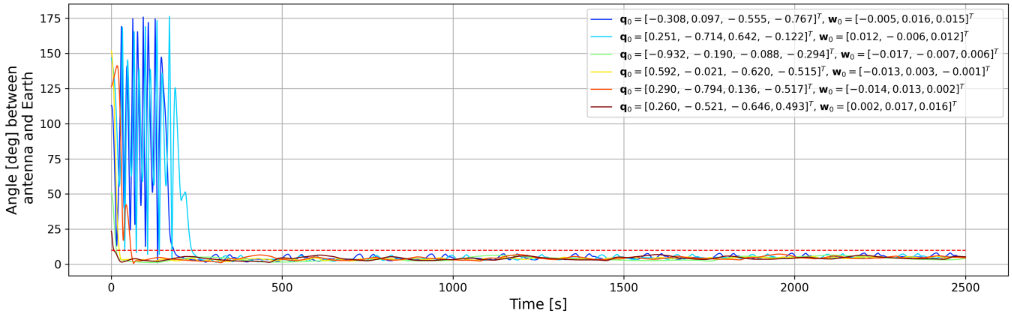}
    }
    \\ 
    \subfigure[$M_z$ Failure Experiment \label{fig:singlefailure(p)-z}]{
        \includegraphics[width=0.7\textwidth]{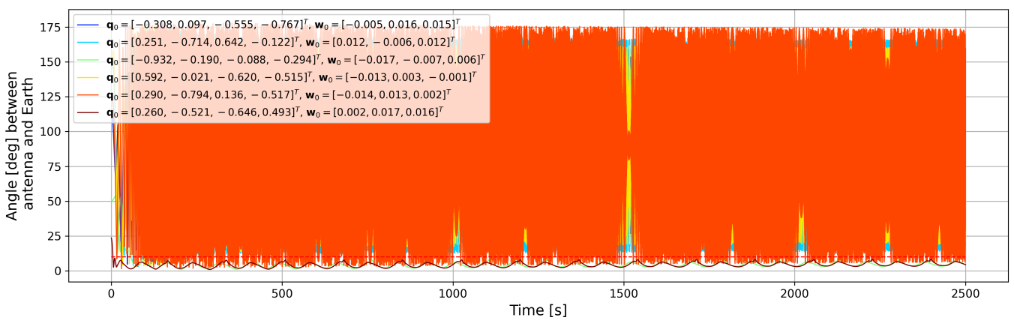}
    }
    \caption{The evaluation of the angle difference $\alpha$ in various initial conditions (Experiment: Single Torque Failure with Previous Reward Function and SAC Algorithm)}
    \label{fig:singlefailure(p)}
\end{figure}

\begin{figure}[htb]
    \centering
    \includegraphics[width=0.7\linewidth]{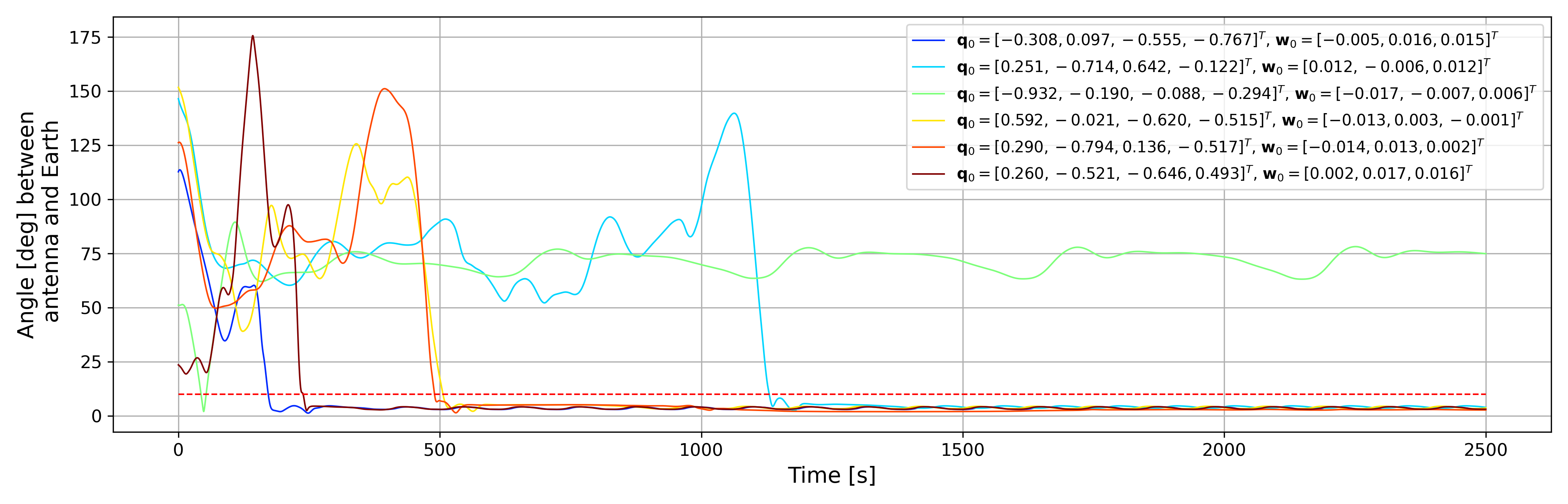}
    \caption{The evaluation of the angle difference $\alpha$ in various initial conditions (Experiment: Torque Noises and Attitude Misalignment with Previous Reward Function and SAC Algorithm)}
    \label{fig:torquenoise+misalignment(p)}
\end{figure}

\begin{figure}[htb]
    \centering
    \subfigure[$M_x$ Failure and Attitude Misalignment Experiment \label{fig:singlefailure+misalignment(p)-x}]{
        \includegraphics[width=0.7\textwidth]{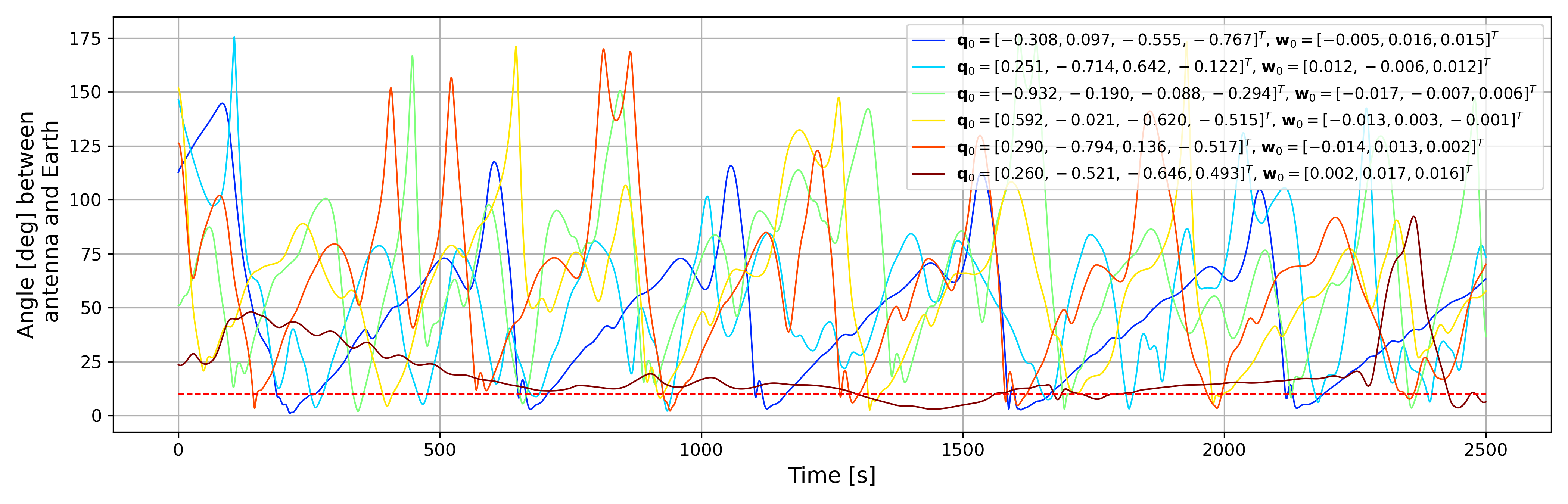}
    }
    \\ 
    \subfigure[$M_y$ Failure and Attitude Misalignment Experiment \label{fig:singlefailure+misalignment(p)-y}]{
        \includegraphics[width=0.7\textwidth]{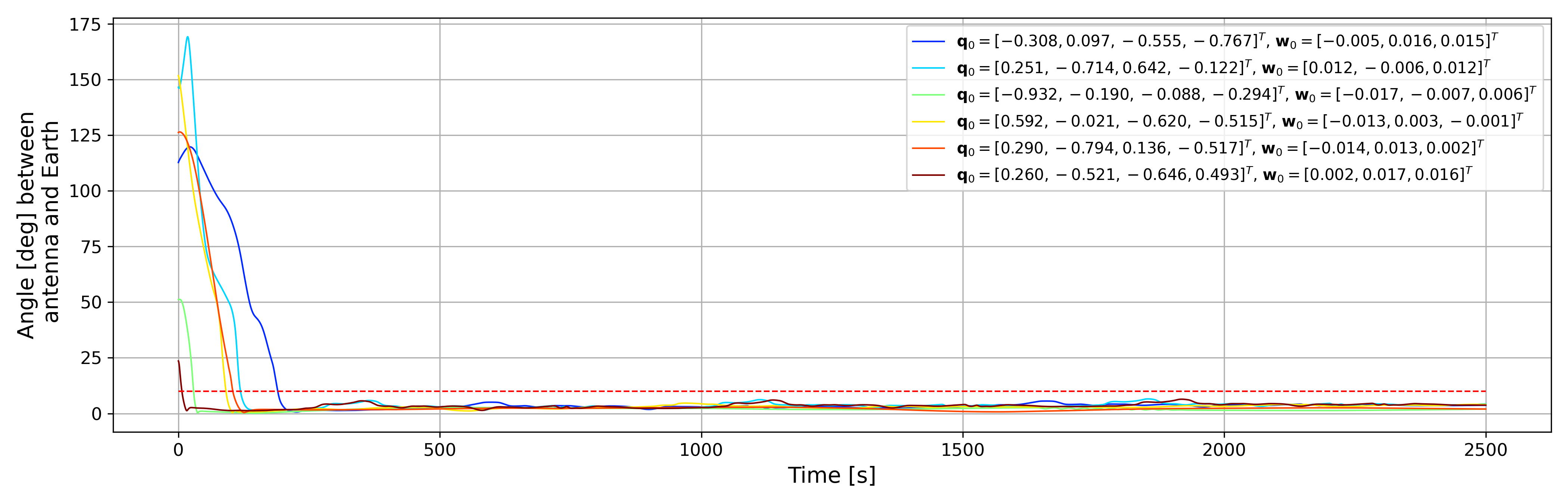}
    }
    \\ 
    \subfigure[$M_z$ Failure and Attitude Misalignment Experiment \label{fig:singlefailure+misalignment(p)-z}]{
        \includegraphics[width=0.7\textwidth]{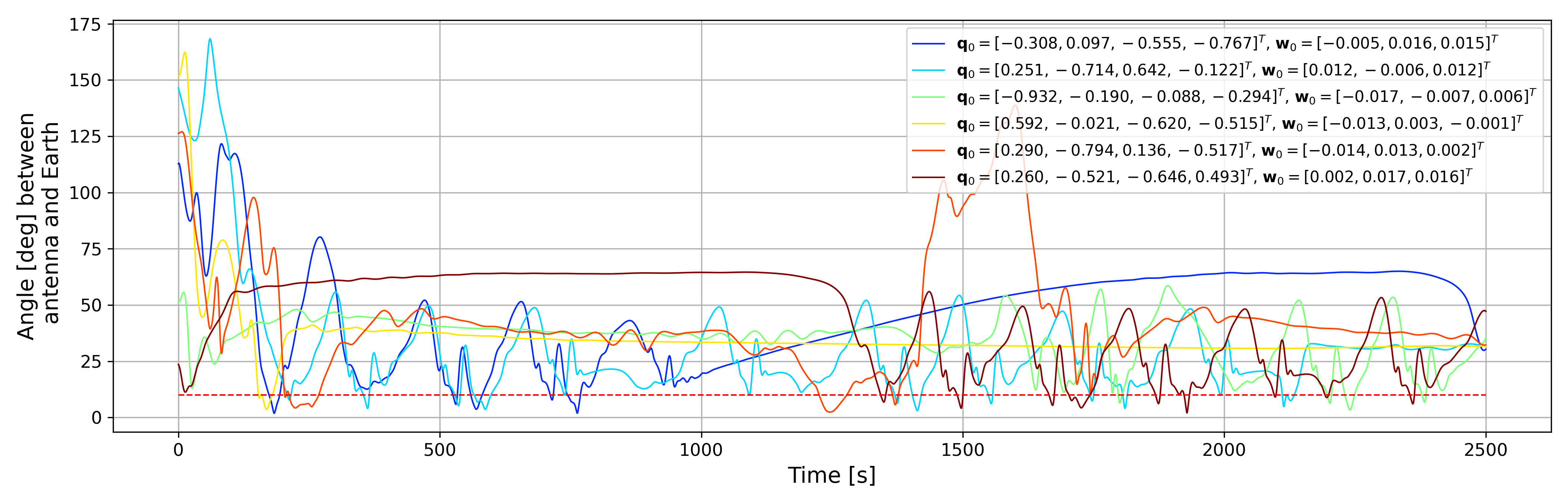}
    }
    \caption{The evaluation of the angle difference $\alpha$ in various initial conditions (Experiment: Single Torque Failure and Constant Attitude Misalignment with Previous Reward Function and SAC Algorithm)}
    \label{fig:singlefailure+misalignment(p)}
\end{figure}

\begin{figure}[htb]
    \centering
    \subfigure[Gyroscope Noise Experiment \label{fig:gyroscopenoise+constant(p)-noise}]{
        \includegraphics[width=0.7\textwidth]{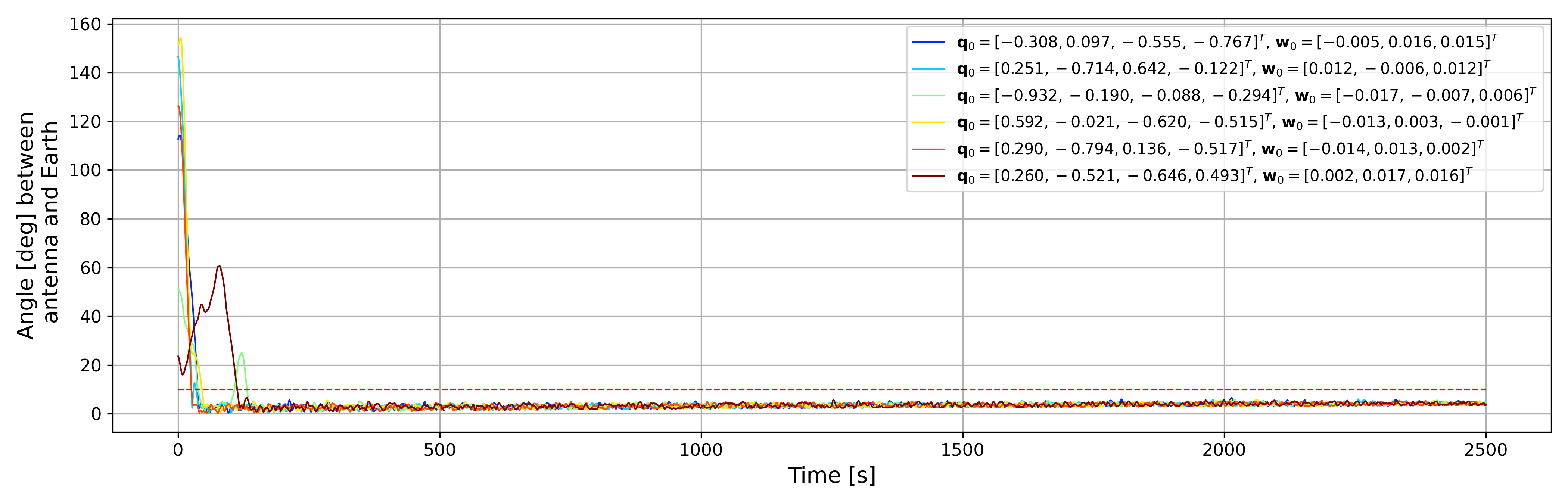}
    }
    \\ 
    \subfigure[Gyroscope Constant Error Experiment \label{fig:gyroscopenoise+constant(p)-constant}]{
        \includegraphics[width=0.7\textwidth]{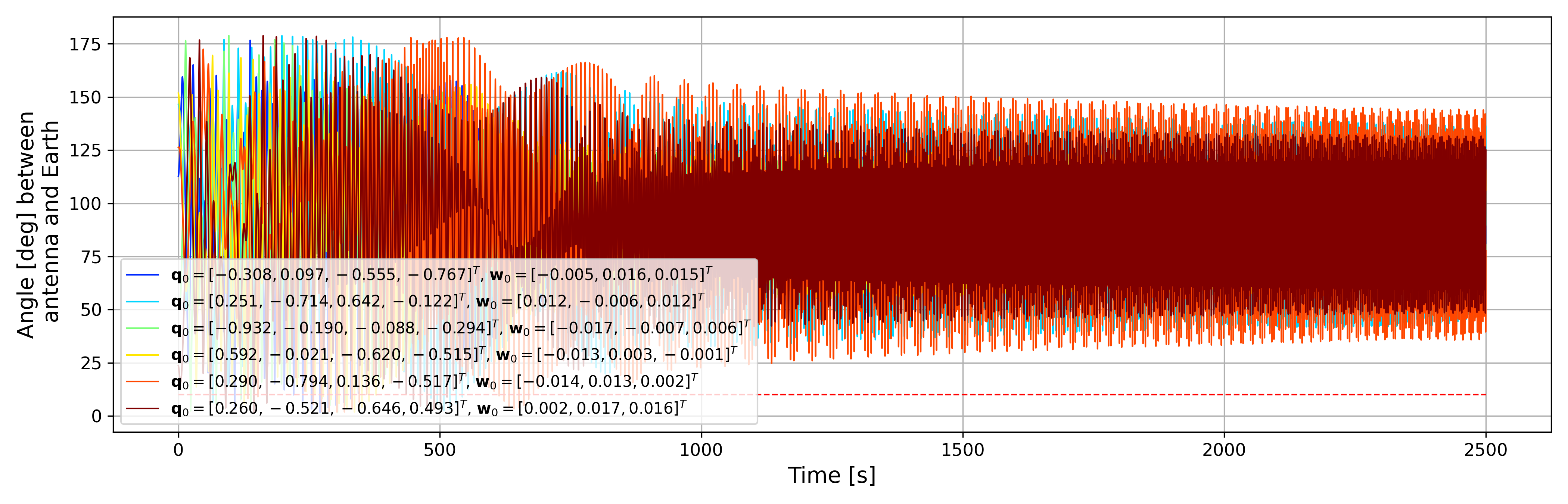}
    }
    \caption{The evaluation of the angle difference $\alpha$ in various initial conditions (Experiment: Gyroscope Errors with Previous Reward Function and SAC Algorithm)}
    \label{fig:gyroscopenoise+constant(p)}
\end{figure}

Here are the results (Figure \ref{fig:baseline(p)}, Figure \ref{fig:singlefailure(p)}, Figure \ref{fig:torquenoise+misalignment(p)}, Figure \ref{fig:singlefailure+misalignment(p)}, and Figure \ref{fig:gyroscopenoise+constant(p)}) with the SAC algorithm and the previous reward function mentioned in Hao et al.'s work \cite{peng2023reorient} .

As we can see in Figure \ref{fig:baseline(p)}, the SAC algorithm can successfully accomplish the control target with the previous reward function. However, results are different while encountering other more difficulty experiments.

Obviously, the agent can't control the satellite in $M_x$ and $M_z$ experiments (shown in Figure \ref{fig:singlefailure(p)-x} and Figure \ref{fig:singlefailure(p)-z}) if we just replace the DDPG algorithm by using the SAC algorithm without adjusting the reward function. For the $M_y$ Failure experiment, refining the reward function of the SAC algorithm has little effect on the experimental results.

Compared with the result shown in Figure \ref{fig:torquenoise+misalignment}, the agent can't ensure that the final angle error of all initial situations falls within the threshold if we don't refine the reward function.

Removing the refined reward function, agents trained by the SAC algorithm only make the error between the satellite antenna and the target direction change periodically in $M_x$ and $M_z$ Failure and Misalignment experiments (shown in Figure \ref{fig:singlefailure+misalignment(p)-x} and Figure \ref{fig:singlefailure+misalignment(p)-z}). Especially for the $M_x$ Failure and Misalignment experiment, the maximum error is significantly greater than the result of SAC training after refining the reward function (shown in Figure \ref{fig:singlefailure+misalignment-a}).

Even though the agent can successfully keep the angular error within the threshold with original reward functions, there are many sawtooth fluctuations in Figure \ref{fig:gyroscopenoise+constant(p)-noise}. Similarly to the agent trained by the SAC algorithm and the refined reward function, the agent over here is still not able to control the satellite while dealing with the gyroscope constant error.

\section{Appendix D: Ablation Study for the GAIL Algorithm}
\label{appendix:AblationLearner}

Some critical hyperparameters of the GAIL algorithm could significantly affect the performance of the learner. After the tuning stage, I select two critical hyperparameters and conduct the ablation study. This study is divided into two parts: For the learner, we substitute the SAC algorithm by using the PPO algorithm. For the expert, we adjust the capacity of the buffer of the expert.

\begin{figure}[htb]
    \centering
    \includegraphics[width=1.0\linewidth]{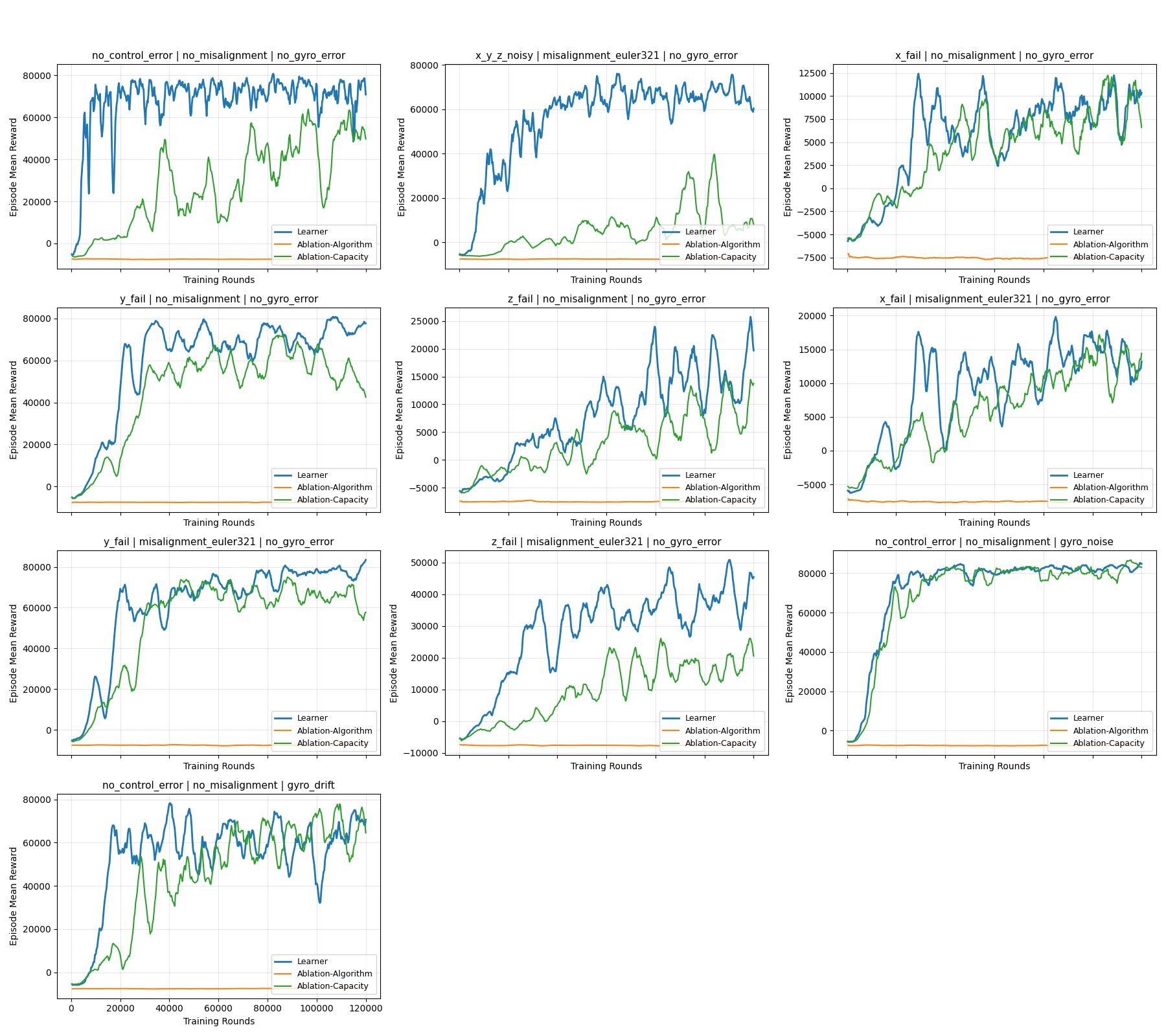}
    \caption{Ablation Study for the GAIL Algorithm}
    \label{fig:AblationLearner}
\end{figure}

As we can see in Figure \ref{fig:AblationLearner}, both of these ablation methods make the performance of the learner worse. If we use the PPO algorithm to train the learner, the learner can't converge and even gets a negative reward value throughout the training. If we decrease the capacity of the buffer of the expert when the learner imitates the expert's behavior, the learner may not be able to gain more experience at each step, which also hinders the learner's training progress.

\bibliographystyle{AAS_publication}   
\bibliography{references}   

\begin{thebibliography}{10}

\bibitem{bolandi2013attitude}
H.~Bolandi, M.~Rezaei, R.~Mohsenipour, H.~Nemati, and S.~M. Smailzadeh, ``Attitude control of a quadrotor with optimized PID controller,''  2013.

\bibitem{lim2001new}
S.~Lim, ``New quaternion feedback control for efficient large angle maneuvers,''  {\em AIAA Guidance, Navigation, and Control Conference and Exhibit}, 2001, p.~4211.

\bibitem{vedant2018dynamic}
V.~Vedant and A.~Ghosh, ``Dynamic programming based attitude trajectories for underactuated control systems,''  {\em 41st Annual AAS Rocky Mountain Section Guidance and Control Conference, 2018}, Univelt Inc., 2018, pp.~191--202.

\bibitem{gao2020satellite}
D.~Gao, H.~Zhang, C.~Li, and X.~Gao, ``Satellite attitude control with deep reinforcement learning,''  {\em 2020 Chinese Automation Congress (CAC)}, IEEE, 2020, pp.~4095--4101.

\bibitem{lei2023adaptive}
J.~Lei, T.~Meng, Y.~Zhu, K.~Wang, and W.~Wang, ``Adaptive Compatible Performance Control for Spacecraft Attitude Control under Motion Constraints with Guaranteed Accuracy,''  05 2023, 10.48550/arXiv.2305.19627.

\bibitem{shi2020disturbance}
K.~Shi, C.~Liu, Z.~Sun, and X.~Yue, ``Disturbance observer-based attitude stabilization for rigid spacecraft with input MRCs,''  {\em Advances in Space Research}, Vol.~66, No.~3, 2020, pp.~689--701.

\bibitem{vedant2019reinforcement}
J.~T. Vedant, ``Reinforcement learning for spacecraft attitude control,''  {\em 70th International Astronautical Congress}, 2019.

\bibitem{elkins2022bridging}
J.~G. Elkins, R.~Sood, and C.~Rumpf, ``Bridging reinforcement learning and online learning for spacecraft attitude control,''  {\em Journal of Aerospace Information Systems}, Vol.~19, No.~1, 2022, pp.~62--69.

\bibitem{peng2023reorient}
H.~Peng and X.~Bai, ``Reorient Satellite Antenna using Reinforcement Learning under Unknown Attitude Failures,''  {\em AIAA SciTech Forum and Exposition, 2023}, American Institute of Aeronautics and Astronautics Inc, AIAA, 2023.

\bibitem{meijkamprobust}
M.~Meijkamp, ``Robust Attitude Control in Active Debris Removal Missions using Reinforcement Learning,''

\bibitem{codevilla2018end}
F.~Codevilla, M.~M{\"u}ller, A.~L{\'o}pez, V.~Koltun, and A.~Dosovitskiy, ``End-to-end driving via conditional imitation learning,''  {\em 2018 IEEE international conference on robotics and automation (ICRA)}, IEEE, 2018, pp.~4693--4700.

\bibitem{ravichandar2020recent}
H.~Ravichandar, A.~S. Polydoros, S.~Chernova, and A.~Billard, ``Recent advances in robot learning from demonstration,''  {\em Annual review of control, robotics, and autonomous systems}, Vol.~3, No.~1, 2020, pp.~297--330.

\bibitem{le2017coordinated}
H.~M. Le, Y.~Yue, P.~Carr, and P.~Lucey, ``Coordinated multi-agent imitation learning,''  {\em International Conference on Machine Learning}, PMLR, 2017, pp.~1995--2003.

\bibitem{ho2016generative}
J.~Ho and S.~Ermon, ``Generative adversarial imitation learning,''  {\em Advances in neural information processing systems}, Vol.~29, 2016.

\bibitem{wiering2012reinforcement}
M.~A. Wiering and M.~Van~Otterlo, ``Reinforcement learning,''  {\em Adaptation, learning, and optimization}, Vol.~12, No.~3, 2012, p.~729.

\bibitem{haarnoja2018soft}
T.~Haarnoja, A.~Zhou, K.~Hartikainen, G.~Tucker, S.~Ha, J.~Tan, V.~Kumar, H.~Zhu, A.~Gupta, P.~Abbeel, {\em et~al.}, ``Soft actor-critic algorithms and applications,''  {\em arXiv preprint arXiv:1812.05905}, 2018.

\bibitem{crassidis2004optimal}
J.~L. Crassidis and J.~L. Junkins, {\em Optimal estimation of dynamic systems}.
\newblock Chapman and Hall/CRC, 2004.

\bibitem{cheetham2021cislunar}
B.~Cheetham, ``Cislunar autonomous positioning system technology operations and navigation experiment (Capstone),''  {\em ASCEND 2021}, p.~4128, 2021.

\bibitem{thienel2003coupled}
J.~Thienel and R.~M. Sanner, ``A coupled nonlinear spacecraft attitude controller and observer with an unknown constant gyro bias and gyro noise,''  {\em IEEE transactions on Automatic Control}, Vol.~48, No.~11, 2003, pp.~2011--2015.

\end{thebibliography}

\end{document}